\documentclass[fleqn,usenatbib]{mnras}

\usepackage{newtxtext,newtxmath}

\usepackage[T1]{fontenc}

\DeclareRobustCommand{\VAN}[3]{#2}
\let\VANthebibliography\thebibliography
\def\thebibliography{\DeclareRobustCommand{\VAN}[3]{##3}\VANthebibliography}

\usepackage{graphicx}	
\usepackage{amsmath}	

\title[MPTA Profile Variability]{Time-Correlated Profile Variability in the MeerKAT Pulsar Timing Array}

\author[B. Bhat et al.]{
Bhavnesh Bhat,$^{1}$\thanks{E-mail:bhavnesh.bhat@postgrad.manchester.ac.uk}
Michael J. Keith,$^{1}$
E. Carli,$^{2,3}$
R.~M.~Shannon$^{2,3}$, and
Benjamin W. Stappers$^{1}$
\\
\small$^{1}$Jodrell Bank Centre for Astrophysics, Department of Physics and Astronomy, The University of Manchester, Manchester M13 9PL, UK\\
$^{2}$Centre for Astrophysics and Supercomputing, Swinburne University of Technology, Hawthorn VIC 3122, Australia\\
$^{3}$OzGrav: The ARC Center of Excellence for Gravitational Wave Discovery, Hawthorn VIC 3122, Australia\\
}

\date{Accepted XXX. Received YYY; in original form ZZZ}

\pubyear{\the\year{}}

\begin{document}
\label{firstpage}
\pagerange{\pageref{firstpage}--\pageref{lastpage}}
\maketitle

\begin{abstract}
Long-term pulse-profile variability has emerged as a potentially important source of uncertainty for high-precision pulsar timing, as changes in the integrated pulse shape can bias pulse times of arrival and affect the sensitivity of pulsar timing arrays to nanohertz gravitational waves. We present a systematic search for long-term pulse-profile variability in the six-year MeerKAT Pulsar Timing Array data, analysing observations of 84 millisecond pulsars using a two-dimensional Gaussian-process framework to identify coherent temporal and phase-dependent profile evolution. We detect pulse-profile variability in 18 pulsars. Using the frequency dependence of the observed profile evolution together with scattering simulations, we classify six pulsars as exhibiting chromatic variability consistent with interstellar scattering, while the remaining twelve display predominantly frequency-independent behaviour suggestive of intrinsic changes in the pulsar emission process. Correcting for the S/N-dependent detection bias, we infer that intrinsic profile variability is substantially more common than implied by the detected sample, with at least 40 per cent of millisecond pulsars exhibiting intrinsic profile variability at the 95 per cent credible level. Motivated by the association between pulse-profile variability and spin-down-rate ($\dot{\nu}$) switching in canonical pulsars, we search for corresponding $\dot{\nu}$ variations but find no statistically significant detections. Simulations indicate that the expected variations remain below the sensitivity of current datasets, although unresolved changes may contribute to weak achromatic timing noise. These results suggest that long-term pulse-profile variability may be common among millisecond pulsars and should be explicitly modelled in future high-precision pulsar-timing experiments.
\end{abstract}
\begin{keywords}
pulsars: general --
methods: data analysis --
scattering 
\end{keywords}



\section{Introduction}
Millisecond pulsars (MSPs)~\citep{Alpar1982,Backer1983} are among the most rotationally stable astrophysical objects known and form the cornerstone of high-precision pulsar timing experiments \citep{Davis1985}. Compared to canonical pulsars, MSPs generally exhibit significantly lower levels of timing noise~\citep{Shannon2010}, enabling exceptionally stable long-term timing solutions. Timing residuals often exhibit stochastic deviations from those predicted by a deterministic timing model~\citep{Helfand1980,Cordes1980}. These deviations can arise from multiple sources, including intrinsic rotational irregularities of the pulsar, propagation effects in the interstellar medium (ISM) \citep{Cordes1984}, and instrumental or calibration systematics \citep{Press1975,Blandford1984}. Intrinsic rotational noise is often characterised as \textit{red noise}, i.e. temporally correlated noise with greater power on long timescales. When such noise is independent of observing frequency, it is commonly referred to as achromatic timing noise.

Radio emission from pulsars exhibits substantial pulse-to-pulse variability~\citep{Lyne1971}, with individual pulses often differing significantly from one rotation to the next. However, when averaged over hundreds or thousands of rotations, pulsars typically produce a remarkably stable integrated pulse profile. MSPs, in particular, possess exceptionally stable time-averaged profiles compared to canonical pulsars~\citep{Helfand1975,Liu2012}, likely reflecting the long-term stability of their magnetospheric emission. This stability underpins precision pulsar timing, where pulse times of arrival (ToAs) can be measured with precisions of the order of microseconds, and in some cases tens of nanoseconds~\citep{Jenet2005,Oslowski2014}. Such precision makes MSPs ideal probes for the detection and characterisation of the stochastic gravitational wave background (GWB)~\citep{foster1990}.

Despite the remarkable stability of their integrated profiles, MSPs can exhibit time-dependent profile variations on both short and long timescales. On short timescales, stochastic pulse-to-pulse fluctuations, commonly referred to as jitter~\citep{Shannon2010,Parthasarathy2021}, arise from variations in the shape and phase of individual pulses and contribute additional uncertainty to the measured ToAs. Other short-timescale phenomena, such as mode changing, have also been reported in MSPs, with variability ranging from individual rotations to timescales of days~\citep{Brook2018,Miles2022,Nathan2023}.

Long-term pulse-profile variability has been studied extensively in canonical pulsars, where changes in the average pulse profile have been shown to correlate with variations in the spin-down rate and contribute to achromatic timing noise~\citep{Lyne2010,Brook2016,Shaw2022,Basu2024,Lower2025,Keith2025}. Understanding whether similar behaviour exists in millisecond pulsars (MSPs) is of particular importance for pulsar timing arrays (PTAs), since even subtle profile changes can bias pulse times of arrival and ultimately limit sensitivity to nanohertz gravitational waves \citep{Lentati2015}.

In contrast to canonical pulsars, long-term profile variability in MSPs has been reported in only a relatively small number of sources. A number of individual MSPs have been studied in detail, such as PSR~J1713$+$0747~\citep{Brook2018,Singha2021,Lin2021,Jennings2024,Mandow2025}, PSR~J0437$-$4715~\citep{Boris2021}, PSR~J1022$+$1001~\citep{Kramer1999,Hotan2004,Liu2015,Shao2016,Feng2021,Padmanabh2021,Fiore2025}, and PSR~J1643$-$1224~\citep{Shannon2016,Brook2016,Boris2021}. 

The first systematic search within a PTA dataset was carried out by \citet{Brook2018}, who identified profile variations in several MSPs and demonstrated that both intrinsic magnetospheric changes and propagation effects through the interstellar medium (ISM) can contribute to the observed variability. These studies have shown that long-term profile evolution in MSPs can occur on timescales ranging from months to years and may arise from either intrinsic changes in the pulsar magnetosphere or propagation effects in the ISM. The frequency dependence of the profile evolution provides one of the most effective diagnostics for distinguishing between these two scenarios.

In canonical pulsars, pulse profile changes are often associated with variations in the spin-down rate~\citep{Lyne2010}, suggesting a connection between magnetospheric state switching and rotational evolution. Whether similar behaviour exists in MSPs remains largely unexplored. Understanding pulse profile stability is therefore essential for precision pulsar timing experiments.

A fundamental assumption in pulsar timing is that the integrated pulse profile remains stable over time. ToAs are typically measured via cross correlation with a de-noised template, and so the ToA can be influenced significantly by small changes in the profile shape~\citep{Taylor1992,Shannon2014}.
Such effects must be carefully characterised and mitigated to prevent contamination or masking of the GWB signal in pulsar timing array data.

In this work, we apply the methodology developed by~\citet{Keith2025} to the MeerKAT Pulsar Timing Array (MPTA) dataset comprising 84 MSPs~\citep{MPTA_data}. Using L-band observations and a two-dimensional Gaussian Process framework, we search for temporal variations in integrated pulse profiles across the MSP population.

Sections~\ref{sec:obs_and_data} and \ref{sec:analysis} describe the observations, data preparation, and analysis pipeline used in this study. Section~\ref{sec:Variable_Pulsar} presents the results of our variability analysis. Section~\ref{sec:Discussions} discusses the implications of these results, and the final section summarises our conclusions.

\section{Observations}
\label{sec:obs_and_data}
The data analysed in this work are drawn from the six-year (2019--2025) MeerKAT Pulsar Timing Array (MPTA) dataset, obtained with the MeerKAT radio telescope~\citep{Jonas2016} in South Africa, operated by the South African Radio Astronomy Observatory (SARAO). The observations span Modified Julian Dates (MJD) 58665--60745 and were conducted using the L-band receiver covering a frequency range of 856--1712\,MHz, centred at 1284\,MHz. Compared to the publicly released 4.5-year MPTA dataset~\citep{MPTA_data}, the dataset analysed here extends the observing baseline by approximately 1.5 years, improving our sensitivity to long-term pulse-profile evolution and enabling the identification of variability on timescales of several years.

The observations were polarisation calibrated prior to recording with the PTUSE backend \citep{Bailes2020} using the procedure described by \citet{Serylak2021}. The MeerKAT polarimetric response has been extensively characterised, demonstrating receptor ellipticities below $1^\circ$ and intrinsic cross-polarisation ratios of $50$--$80\,\mathrm{dB}$ across the observing band \citep{Serylak2021}. Since the residual calibration errors are expected to vary between observing epochs, any remaining instrumental artefacts should primarily introduce stochastic fluctuations in the pulse profile rather than the coherent, smoothly evolving structures observed in our Gaussian-process residual maps.

Radio-frequency interference (RFI) mitigation and initial data processing were performed using the automated MeerTime processing pipeline (\textsc{MEERPIPE}\footnote{\href{https://github.com/nf-core/meerpipe.git}{https://github.com/nf-core/meerpipe}}), which incorporates the \textsc{MEERGUARD}\footnote{\href{https://github.com/danielreardon/MeerGuard.git}{https://github.com/danielreardon/MeerGuard}} algorithm for robust RFI excision. All data products are archived on the OzStar supercomputing facility at Swinburne University of Technology.

The MPTA monitored a larger sample of pulsars, of which 84 were included in this analysis after excluding sources with insufficient timing precision. The remaining pulsars are observed on an approximately bi-weekly cadence. Observations were conducted per epoch at L-band for uniformity across the dataset with maximum observation duration 2048 s, with integration times dependent on individual pulsars to optimize timing precision while maintaining a high signal-to-noise ratio (S/N). This approach allowed a large sample of pulsars to be monitored within the MPTA sample set \citep{MPTA_data}. The data were coherently de-dispersed in real time at a constant dispersion measure (DM) and folded at the predicted topocentric spin period of the pulsar.

The observations are stored in PSRFITS~\citep{Hotan2004} format archive files, containing full-Stokes (I, Q, U, V) pulse profiles with 8-s subintegrations and 1024 frequency sub-bands across the observing band.
For the analysis presented here, the data are frequency-averaged to 32 frequency sub-bands and time-scrunched to a single integrated profile per observation to boost signal-to-noise. 

\section{Analysis}
\label{sec:analysis}
 Each of the 32 frequency subbands is individually aligned in pulse phase and amplitude-scaled using frequency-resolved templates (``portraits'') constructed with the \textsc{PulsePortraiture} software~\citep{Pennucci2014,Pennucci2019}. The alignment is performed using the Fourier-domain template-matching algorithm of \citet{Taylor1992}, implemented in the \textsc{ProfileShiftFit} routine of \textsc{PSRCHIVE}~\citep{Hotan2004,VanStraten2006}. For each observing epoch, an independent phase offset and amplitude scale are fitted in each frequency subband relative to the corresponding frequency-resolved template. The amplitude scaling removes sensitivity to epoch-to-epoch variations in total flux density arising from interstellar scintillation, instrumental gain variations, and residual calibration uncertainties, allowing the subsequent analysis to focus on changes in the pulse-profile morphology rather than its absolute flux density. Performing this alignment and scaling prior to frequency averaging also reduces the impact of small dispersion-measure (DM) variations and other chromatic propagation effects, which could otherwise introduce frequency-dependent distortions into the averaged pulse profile.

Observations with a total profile signal-to-noise ratio (S/N), measured from the frequency-integrated pulse profile, of lower than 10 are excluded to prevent low-quality data from introducing excess noise and bias into the profile residuals and subsequent 2-D Gaussian-process analysis. Following alignment and quality selection, the 32 frequency sub-bands are averaged to form a single frequency-integrated pulse profile, and only the total intensity (Stokes~I) is retained for the subsequent analysis.

\subsection{Data Preparation}
The average pulse profiles per observation are arranged as a two-dimensional stack, with axes corresponding to the rotational phase of the pulsar and the observational epoch. Owing to the periodic nature of the pulse signal, this representation can be equivalently viewed as a one-dimensional time series by treating phase bins as sequential samples within each rotation.

For the analysis, we restrict the data to the on-pulse region, which is defined following the procedure described by \citet{Keith2025}. Phase bins outside this window are assumed to contain only radiometer noise and are excluded from the Gaussian-process modelling, as their inclusion would introduce additional variance and potentially bias the inferred phase-dependent variability.

A median profile computed across all epochs is first subtracted from each individual observation. After correcting the baseline of each residual profile, an inverse-noise-variance-weighted mean residual profile is also removed, where the weights are given by the inverse square of the off-pulse rms for each observation. The resulting residual profiles are used to construct the residual map (middle panel of the lower row in Figure~\ref{fig:J1713}). This highlights deviations from the average profile, with temporally correlated structures across pulse phase often visible in the resulting heat map.

To further characterise these variations, we employ the \textsc{psrcelery} framework~\citep{Keith2025}, which models the data using a two-dimensional Gaussian process, enabling the detection and quantification of subtle, time-dependent profile variability.

\subsection{Two Dimensional Gaussian Process}
The kernel used in this analysis is a two-dimensional model constructed as the product of two one-dimensional kernels, describing correlations along the pulse phase and observational epoch axes. This formulation allows the Gaussian process (GP) to capture variations in the pulse profile as a function of both phase and time. 

The covariance kernel adopted in this analysis is constructed as the product of two components: a phase kernel and a time kernel. The phase kernel characterises correlations between different pulse-phase bins within an individual pulse profile and is periodic, with a period equal to one rotation of the pulsar (i.e. one pulse phase cycle). In contrast, the time kernel describes correlations between pulse profiles observed at different epochs.

In this work, we adopt the simple model provided in \textsc{psrcelery}, where the phase dependence is modelled using a Gaussian kernel, while the temporal correlation is described using a single-term \textsc{celerite} kernel, namely the Matérn-$3/2$ kernel~\citep{Foreman-Mackey2017}, characterised by an amplitude $A$ and a characteristic timescale $\lambda$. The kernel is given by
\begin{equation}
    k(\tau) = A\left(1+\frac{\sqrt{3}\tau}{\lambda}\right)\exp\left(-\frac{\sqrt{3}\tau}{\lambda}\right),
\end{equation}
where $\tau = |t_i - t_j|$ is the time separation between two observations, $A$ is the covariance amplitude, and $\lambda$ is the characteristic correlation timescale over which the profile variations remain coherent. 

We adopt the Matérn-$3/2$ temporal kernel following previous GP analyses of long-term pulse-profile variability \citep{Brook2016,Shaw2022,Lower2025,Keith2025}, where Matérn kernels have been found to provide a useful description of the observed time-correlated profile variations. There is, however, no unique choice of temporal kernel for this problem. \textsc{psrcelery} allows other single-term \textsc{celerite} kernels to be used, while more complicated kernels can also be constructed at additional computational cost \citep{Keith2025}.
The two-dimensional Gaussian process provides a continuous reconstruction of the pulse-profile evolution in both pulse phase and time, enhancing the visibility of coherent low-level profile variations that may be difficult to identify directly in the residual maps.

\section{Results}
\label{sec:Variable_Pulsar}
We find long-term (i.e. greater than one observation) profile variations in a total of 18 pulsars. Candidate variable pulsars were initially identified as those showing significant, coherent structure in the Gaussian-process reconstruction, and were subsequently verified by direct inspection of the profile residuals to confirm that the apparent variations correspond to genuine time-correlated profile changes rather than isolated fluctuations consistent with noise. These pulsars fall broadly into three categories.
Firstly, and perhaps of most interest, are PSRs J0900$-$3144, J1103$-$5403, and J2124$-$3358, which show quasi-periodic profile-shape variations. Whilst the individual timescales, shapes, and properties differ from pulsar to pulsar, they are overall very reminiscent of those in the \citet{Lyne2010} sample of canonical pulsars.
Secondly, there are pulsars such as PSR~J1713+0747, which exhibit discrete profile-change events followed by a slow evolution towards their pre-event profile, although the profile does not fully recover over the current observing baseline. These may, of course, also be analogous to the state switching observed in canonical pulsars, where individual emission states have been observed to persist for more than two decades. The phenomenology of these transitions is remarkably diverse, ranging from abrupt changes between discrete states to smooth, gradual evolution of the pulse profile.
The final category comprises sources that show a more stochastic process, where the profile evolves slowly over timescales of years. With further investigation, we find that this behaviour is associated with strong spectral evolution (i.e. more significant at lower frequencies), and that the observed profile changes are broadly consistent with the convolution of the profile with an exponential tail, strongly suggesting that the variability is associated with changes in pulse broadening due to interstellar scattering.
Here we focus on the first two classes, addressing each pulsar in turn. Further discussion of profile variations consistent with scattering is given in Appendix~\ref{sec:Scattering_cases}. We also present in Appendix~\ref{sec:other_intr_pulsar} a number of candidate intrinsically variable pulsars whose profile changes are suggestive but too weak to establish conclusively with the current observations.

\subsection{PSR J1713+0747}
PSR~J1713$+$0747 is one of the most precisely timed pulsars in current pulsar timing arrays. Three pulse-profile change events have been reported in this pulsar. Two relatively small-amplitude events were initially suggested to arise from interstellar propagation effects~\citep{Brook2018}; however, subsequent analyses favoured a magnetospheric origin for these events~\citep{Boris2021}. A third, much larger profile-change event occurred near the middle of our observing span and has been the subject of extensive follow-up studies~\citep{Singha2021,Lin2021,Jennings2024}.

\begin{figure*}
    \centering
    \includegraphics[width=0.86\linewidth]{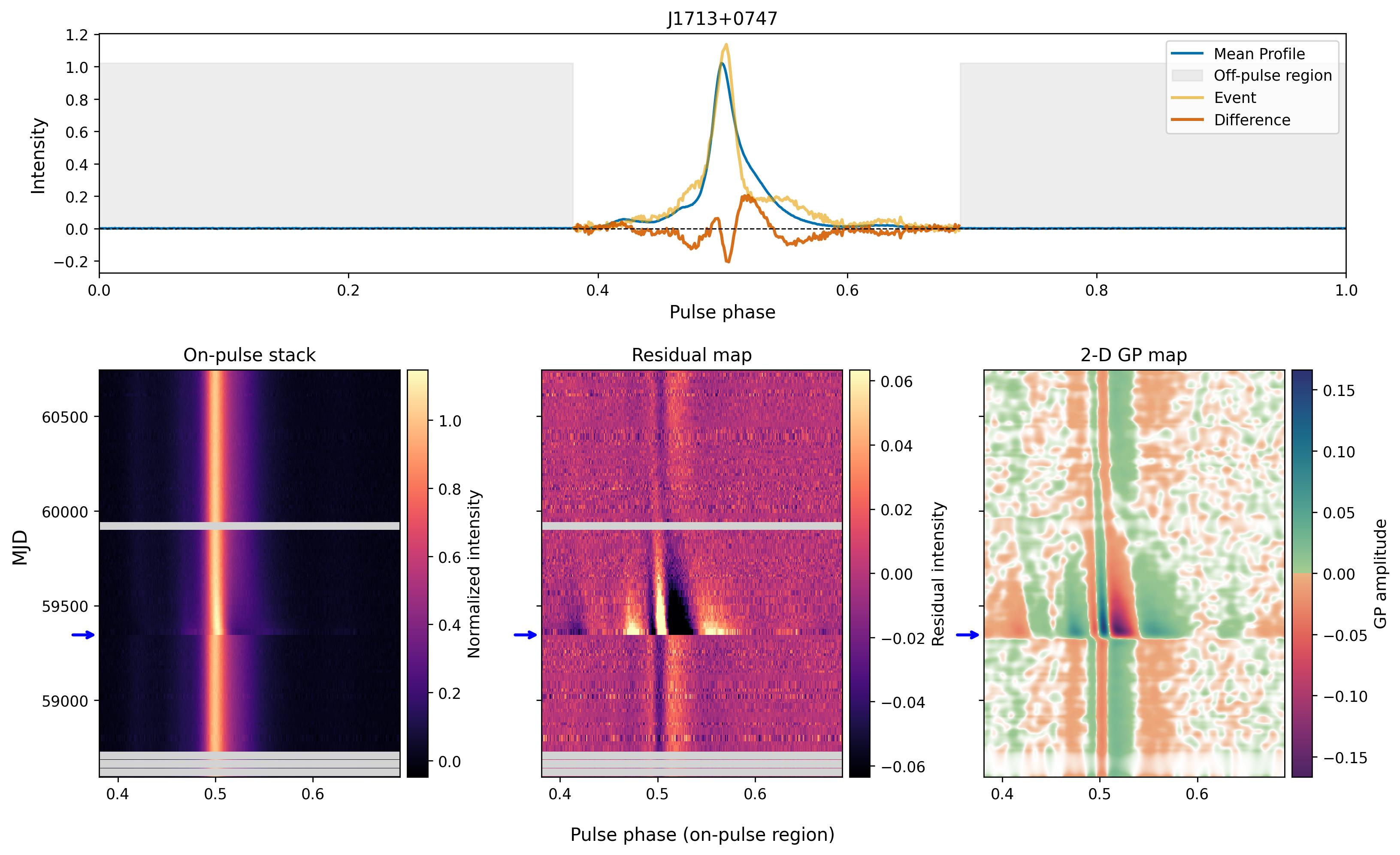}
    \caption{Pulse-profile variability of PSR~J1713$+$0747. \textit{Top:} The mean pulse profile is shown in blue, the profile during the variability event is shown in gold, and the difference between the two profiles is shown in dark orange. The difference profile has been scaled by a factor of five for clarity. The grey shaded regions indicate the off-pulse phase, which is excluded from the variability analysis. \textit{Bottom left:} Stack of the on-pulse profiles as a function of observing epoch, with the colour scale representing the normalised pulse intensity. \textit{Bottom centre:} Residual map obtained after subtracting the weighted mean profile from each observation, where the colour scale represents the residual intensity. \textit{Bottom right:} Two-dimensional Gaussian-process reconstruction of the profile variability, with the colour scale indicating the reconstructed GP amplitude. Grey horizontal bands in the lower panels indicate extended intervals with no observations, for which no profile information is available and hence no interpolation is shown. The blue arrows indicate the onset of the well-known profile-change event near MJD~59320, which evolves to a maximum amplitude before recovering approximately exponentially.}
    \label{fig:J1713}
\end{figure*}

In our analysis we clearly detect this highly significant event, as shown in Figure~\ref{fig:J1713}.
The initial sudden profile change at MJD~59320 is followed by an exponential recovery of the pulse shape, eventually returning to the pre-event shape.
We do not discuss this event further as very detailed studies already exist~\citep{Mandow2025}.

\subsection{PSR J0900$-$3144}
\begin{figure*}
    \centering
    \includegraphics[width=0.86\linewidth]{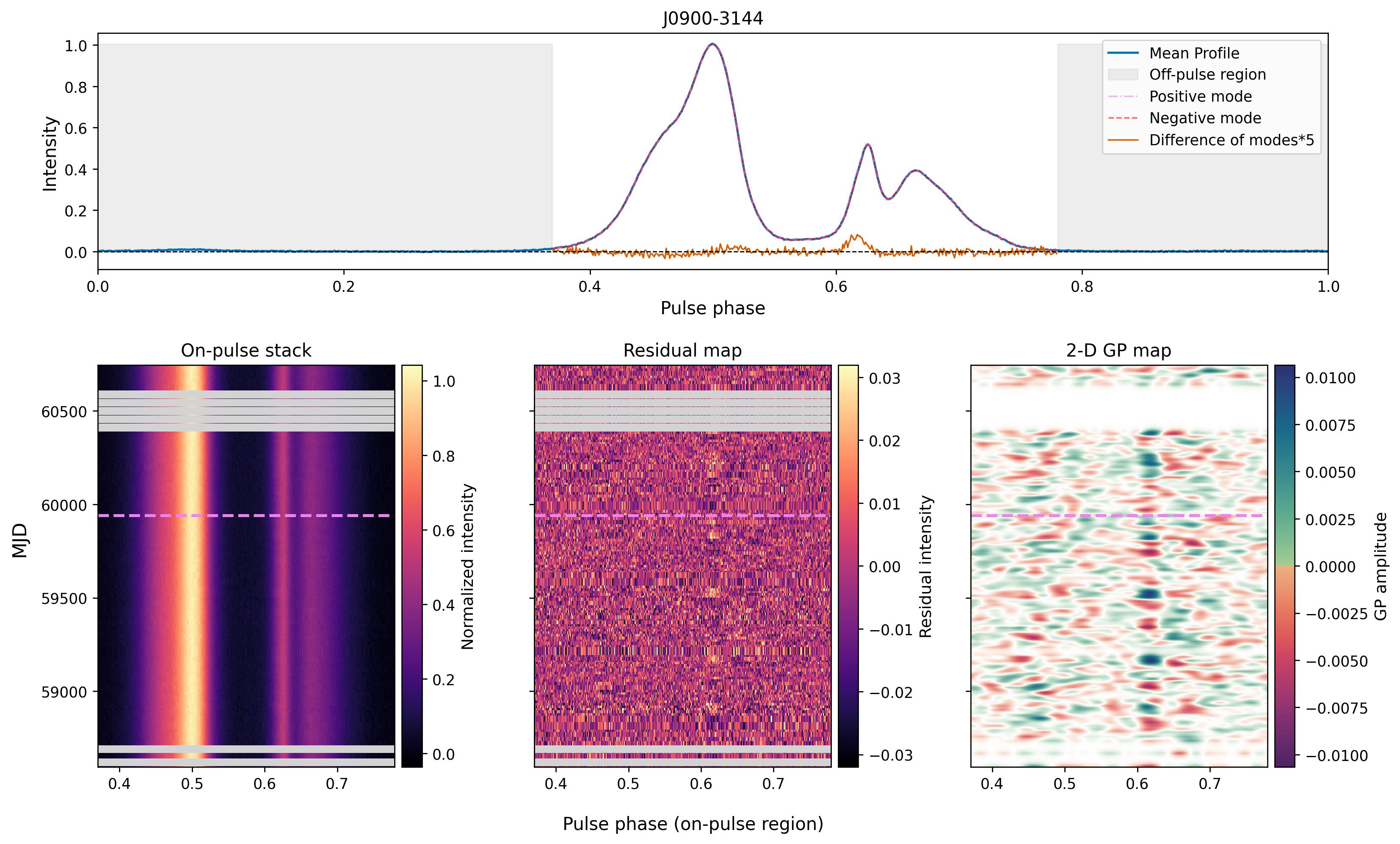}
    \caption{As Figure~\ref{fig:J1713}, but for PSR~J0900$-$3144. The top panel shows representative profiles corresponding to the positive and negative deviations from the median profile, determined from the variation in the most variable pulse-phase region in their difference profile. The quasi-periodic profile variability is primarily confined to the trailing component of the pulse profile. The dashed horizontal line marks the epoch of the glitch demonstrating that no obvious change in the profile variability is associated with the event.}
    \label{fig:J0900}
\end{figure*}

The variability in the profile of this pulsar appears as a quasi-periodic variation of the leading shoulder of the second component, with an amplitude of about 2 per cent of the peak (see Figure~\ref{fig:J0900}).
The quoted profile-change percentage is defined as the ratio of the maximum absolute difference between the two profiles to the peak intensity of the mean pulse profile.
We estimate the period of the variations to be $\sim350\pm50$\,d using a Lomb--Scargle periodogram~\citep{Lomb1976,Scargle1982} of the Profile Change Time Series (PCTS), obtained by projecting the 2-D Gaussian-process reconstruction onto its first principal component.
Although this timescale is close to one year, we find no clear correspondence with either the solar-wind geometry or the measured DM variations.

This pulsar experienced a glitch at MJD~59942~\citep{Bhat2026}, corresponding to a sudden increase in its rotational frequency, which is thought to arise from internal disturbances within the neutron star. Although pulse-profile changes have been reported in association with glitches in some pulsars~\citep{Weltevrede2011,Keith2013}, we detect no evidence that the profile variability is related to the glitch.

\subsection{PSR J1103$-$5403}
\begin{figure*}
    \centering
    \includegraphics[width=0.86\linewidth]{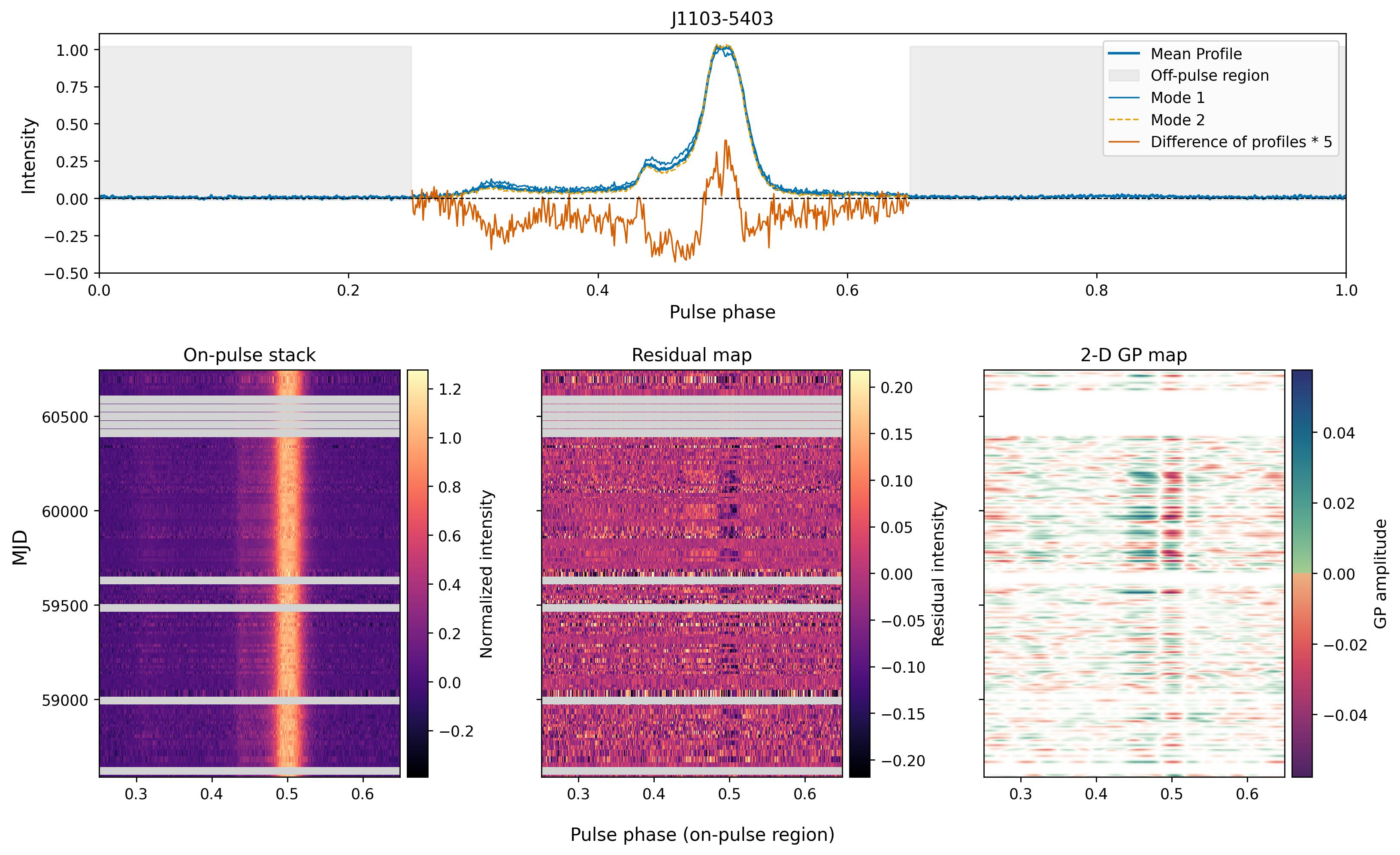}
    \caption{As Figure~\ref{fig:J1713}, but for PSR~J1103$-$5403. The top panel shows the two representative profile modes identified from the method in Figure~\ref{fig:J0900} with their difference profile. The variability is characterised by mode switching, with the most significant changes occurring on the leading edge and at the peak of the main pulse.}
    \label{fig:J1103}
\end{figure*}

PSR~J1103$-$5403 was previously included in the MPTA dataset but has since been identified to exhibit significant quasi-periodic profile variations~\citep{Nathan2023}.
Owing to these pronounced profile changes it was subsequently excluded from later MPTA analyses due to the resulting excess noise~\citep{MPTA_data}. 
Our analysis (Figure~\ref{fig:J1103}) clearly shows the two distinct emission states seen by \cite{Nathan2023}, which are characterised by a slight change in the relative intensity of the narrow main peak and the broader underlying component.
A Lomb--Scargle periodogram applied to the PCTS yielded a dominant period of $\sim103 \pm 56$\,d, providing an initial estimate of the characteristic recurrence timescale. Given the quasi-periodic nature of the pulse-profile variability, this period is not interpreted as evidence for a strictly coherent periodic signal, but rather as an approximate recurrence timescale.

\subsection{PSR J1525$-$5545}
\begin{figure*}
    \centering
    \includegraphics[width=0.86\linewidth]{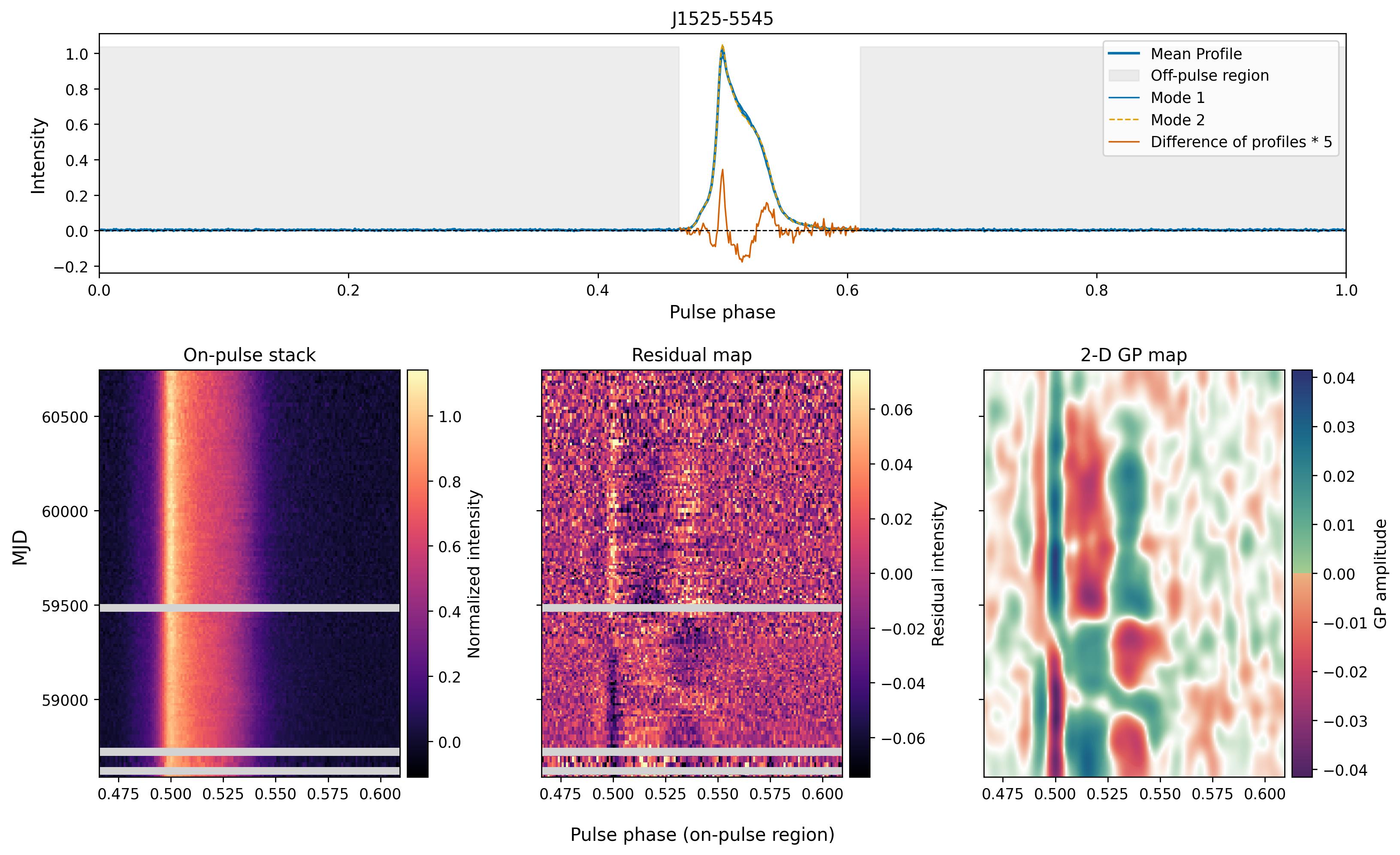}
    \caption{As Figure \ref{fig:J1713} but for J1525$-$5545. The variability is concentrated around the leading edge and peak of the main pulse, where a localized intensity change produces a coherent feature in the Gaussian-process reconstruction.}
    \label{fig:J1525}
\end{figure*}

The profile of this pulsar shows a single change in the intensity of the narrow peak of the profile, which seems at least partly correlated with a change in the trailing part of the profile (see Figure~\ref{fig:J1525}).
In general, when the next leading peak is strong, the trailing component is also strong, and the middle of the profile is suppressed, however, around MJD 59000 the leading component is weak at the same time as a strong trailing component.
Similar behaviour has been reported in canonical pulsars, with PSR~B0740$-$28 representing an extreme example in which the correlations between profile components invert over time (\citealt{Keith2025}). The overall evolution also resembles the complex ``swooshing'' behaviour observed in PSR~J2043+2740, where multiple profile components evolve differently and cannot be fully described by a single mode of variability (\citealt{Keith2025}).

With a relatively high DM of approximately \(126\,{\rm pc\,cm^{-3}}\) \citep{Cherry2014} and known steep spectrum chromatic variations reported in the timing analysis~\citep{MPTA_data}, we initially suspected these changes may be due to the interstellar medium.
However, the observed profile variation does not show a strong frequency dependence in our analysis, and the profile shape changes are hard to reproduce by convolution with a scattering-like impulse response function.
Therefore we argue that these are most likely to be intrinsic profile variations.

\subsection{PSR J1547$-$5709}
\begin{figure*}
    \centering
    \includegraphics[width=0.86\linewidth]{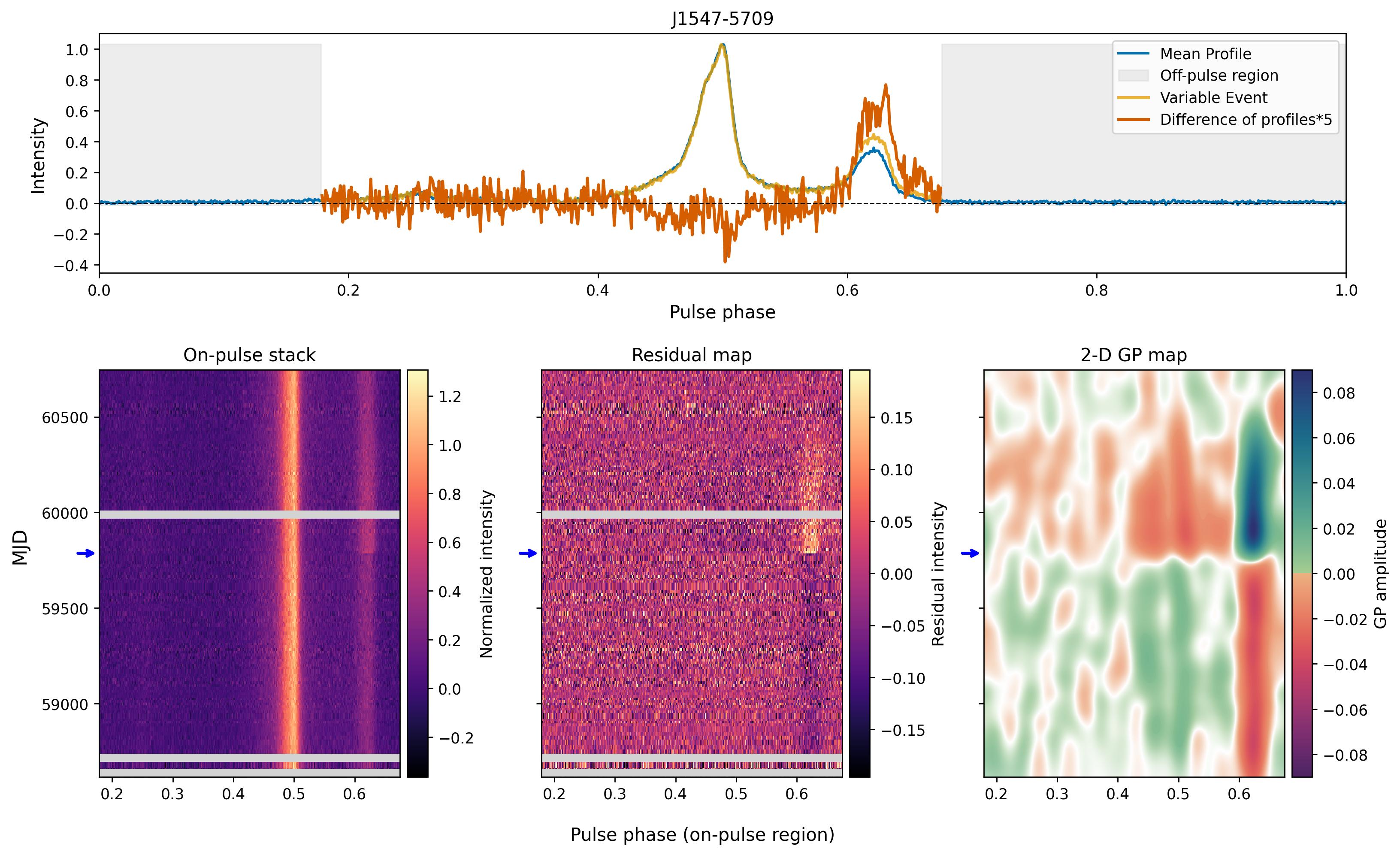}
    \caption{As Figure~\ref{fig:J1713}, but for PSR~J1547$-$5709. The top panel shows the profile during the transient event. The blue arrows indicate the onset of a discrete profile-change event, during which the peak amplitude of the second pulse component increases before relaxing approximately exponentially back to its pre-event state.}
    \label{fig:J1547}
\end{figure*}

The pulse-profile evolution of this pulsar is characterised by a one-off brightening of the second pulse component~\textcolor{red}{\citep{Wei2026}}. 
The peak amplitude of this component increases by approximately $15$ per cent around MJD~59786.3 (Figure~\ref{fig:J1547}), before recovering approximately exponentially to its pre-event level. 
The same behaviour is clearly visible in the residual map and the 2-D Gaussian-process reconstruction, which isolate the coherent temporal evolution of the event.

\subsection{PSR J1721$-$2457}
\begin{figure*}
    \centering
    \includegraphics[width=0.86\linewidth]{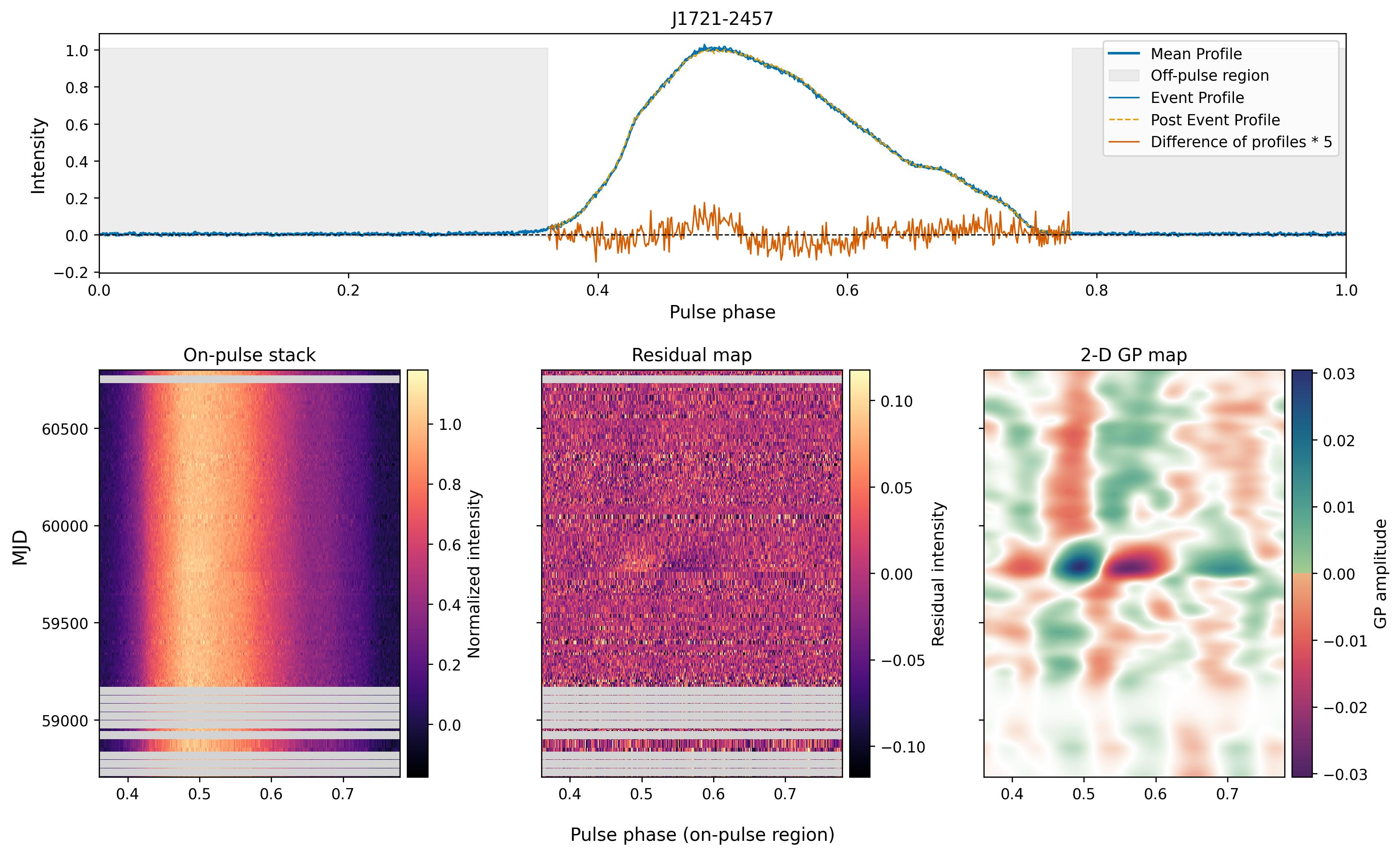}
    \caption{As Figure~\ref{fig:J1713}, but for PSR~J1721$-$2457. A localized profile-change event is observed, primarily affecting the central and trailing regions of the pulse profile. While the event is only weakly discernible in the residual map, it is clearly isolated by the 2-D Gaussian-process reconstruction.}
    \label{fig:J1721}
\end{figure*}

This pulsar exhibits a small but significant profile shape variation between MJD~59608.15 and 60056.04, as shown in Figure~\ref{fig:J1721}. The variability is primarily concentrated near the central and trailing regions of the generally featureless broad pulse profile.
The profile-change event is centred around MJD 59814.6 and is characterised by a coherent redistribution of pulse intensity across the profile. 
Following the event, the pulse profile does not appear to recover immediately to its pre-event morphology. Instead, the 2-D Gaussian-process reconstruction indicates that the post-event profile remains systematically different from the pre-event state throughout the remainder of the dataset.

\subsection{PSR J2124$-$3358}
PSR~J2124$-$3358 exhibits quasi-periodic variations with amplitude around 2 per cent over a narrow pulse phase window centred on the main peak of the pulsar.
A Lomb-Scargle periodogram analysis on the PCTS yields a characteristic period of $652\pm197$ days. 
\begin{figure*}
    \centering
    \includegraphics[width=0.86\linewidth]{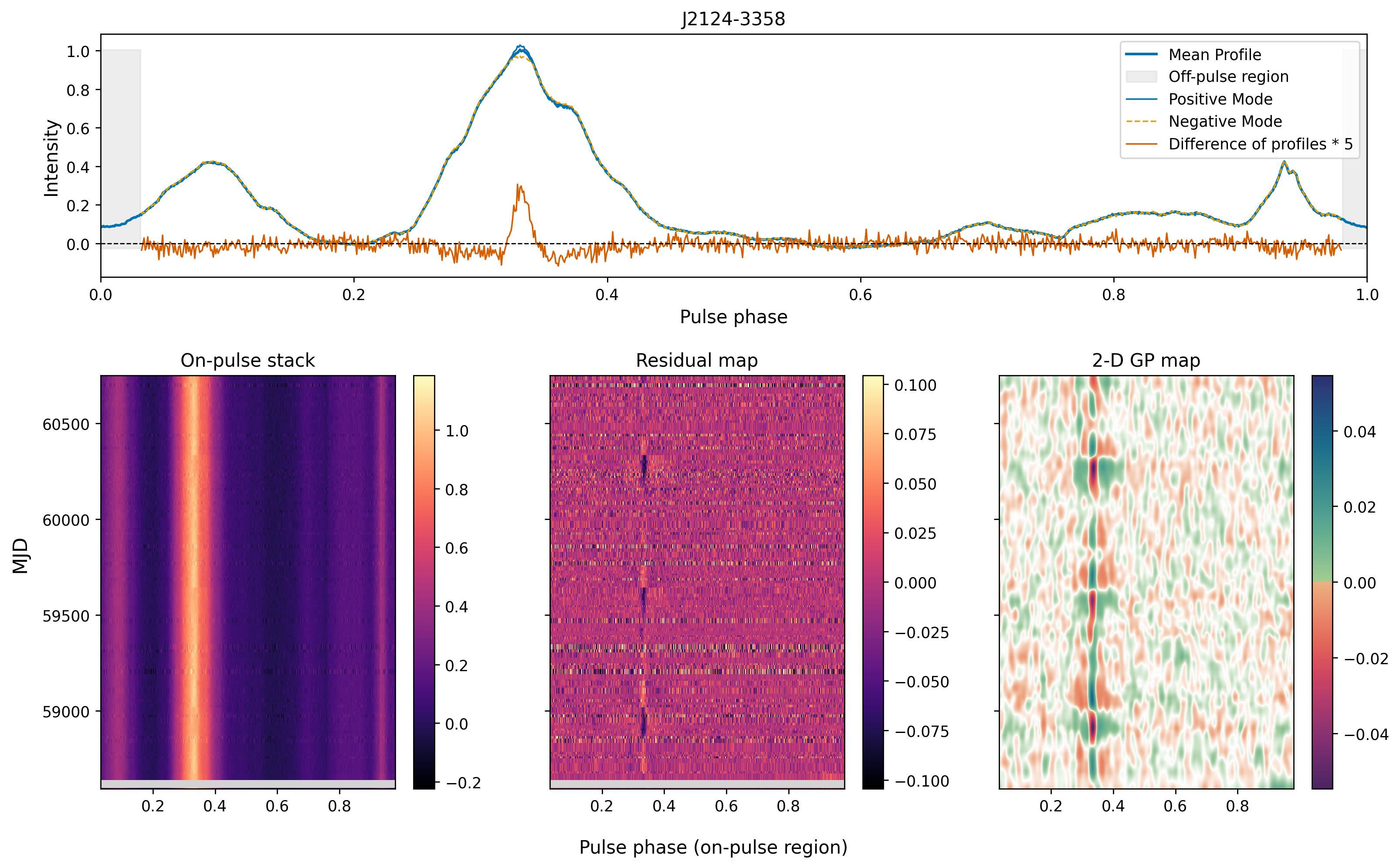}
    \caption{As Figure \ref{fig:J1713} but for PSR~J2124$-$3358. The variability is primarily localised to the central pulse component and exhibits quasi-periodic behaviour, with the Gaussian-process reconstruction emphasising the coherent temporal evolution of the profile change.}
    \label{fig:J2124}
\end{figure*}

\subsection{PSR J0437$-$4715}
\begin{figure*}
    \centering
    \includegraphics[width=0.86\linewidth]{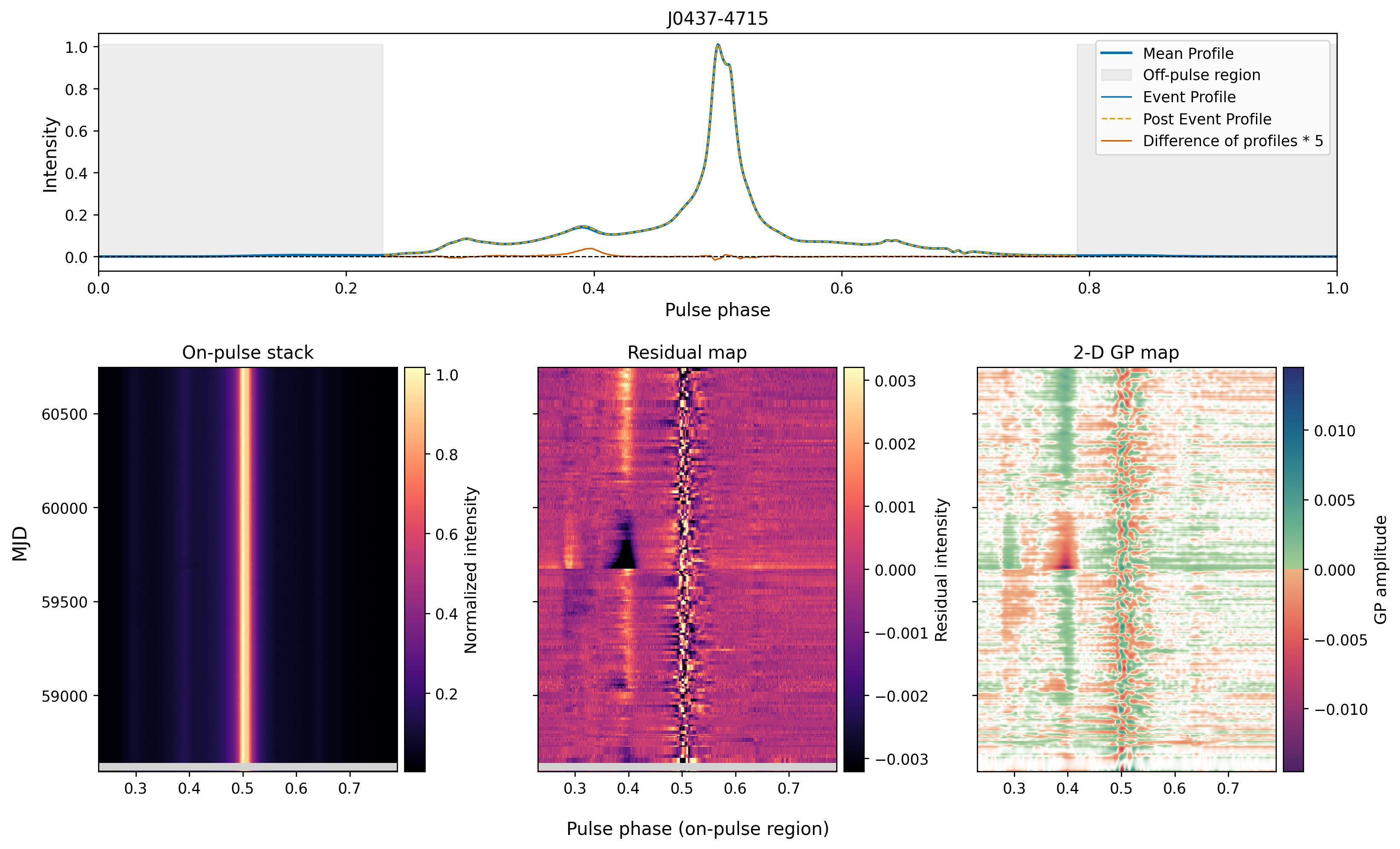}
    \caption{As Figure \ref{fig:J1713} but for PSR~J0437$-$4715. The Gaussian-process model highlights a coherent profile-change event together with long-term low-level quasi-periodic variability, consistent with the complex profile evolution reported for this pulsar.}
    \label{fig:J0437}
\end{figure*}

PSR~J0437$-$4715 has also been the subject of numerous studies investigating pulse-profile stability, polarisation calibration, and timing systematics (e.g. \citealt{Manchester1995,Navarro1997,Oslowski2014}). This pulsar is known to exhibit significant achromatic timing noise and has previously shown a pulse-profile variation event around MJD~57070 \citep{Boris2021}. In our analysis, we identify a similar transient event around MJD~59670 in Figure~\ref{fig:J0437}, characterised by a sudden profile change followed by an approximately exponential recovery. This raises the possibility that such events may recur in PSR~J0437$-$4715.

The transient event corresponds to a measured profile change of approximately 1.4 per cent. Given the exceptional timing precision of PSR~J0437$-$4715, even such small profile perturbations may have measurable effects on timing residuals. A more detailed analysis of the profile variability and its impact on timing will be presented in Mandow et al. (in prep.) using data from both Parkes and MeerKAT.

In addition to the transient event, we also find evidence for persistent stochastic variability concentrated near the peak of the main pulse. Previous single-pulse studies have shown that this phase region is associated with strong pulse-to-pulse intensity fluctuations, orthogonally polarised modes, and stochastic wideband impulse-modulated self-noise (SWIMS), all of which can contribute to pulse jitter and limit timing precision \citep{Oslowski2014}. The phase localisation of the variability observed here therefore suggests that at least part of the stochastic profile variation may arise from pulse jitter noise rather than long-term changes in the emission process itself.

\subsection{PSR J1022+1001}
PSR~J1022$+$1001 is well known for exhibiting significant pulse-profile variability and has been extensively studied in the literature~\citep{Kramer1999,Hotan2004,Liu2015,Shao2016,Feng2021,Padmanabh2021,Fiore2025}.

\begin{figure*}
    \centering
    \includegraphics[width=0.86\linewidth]{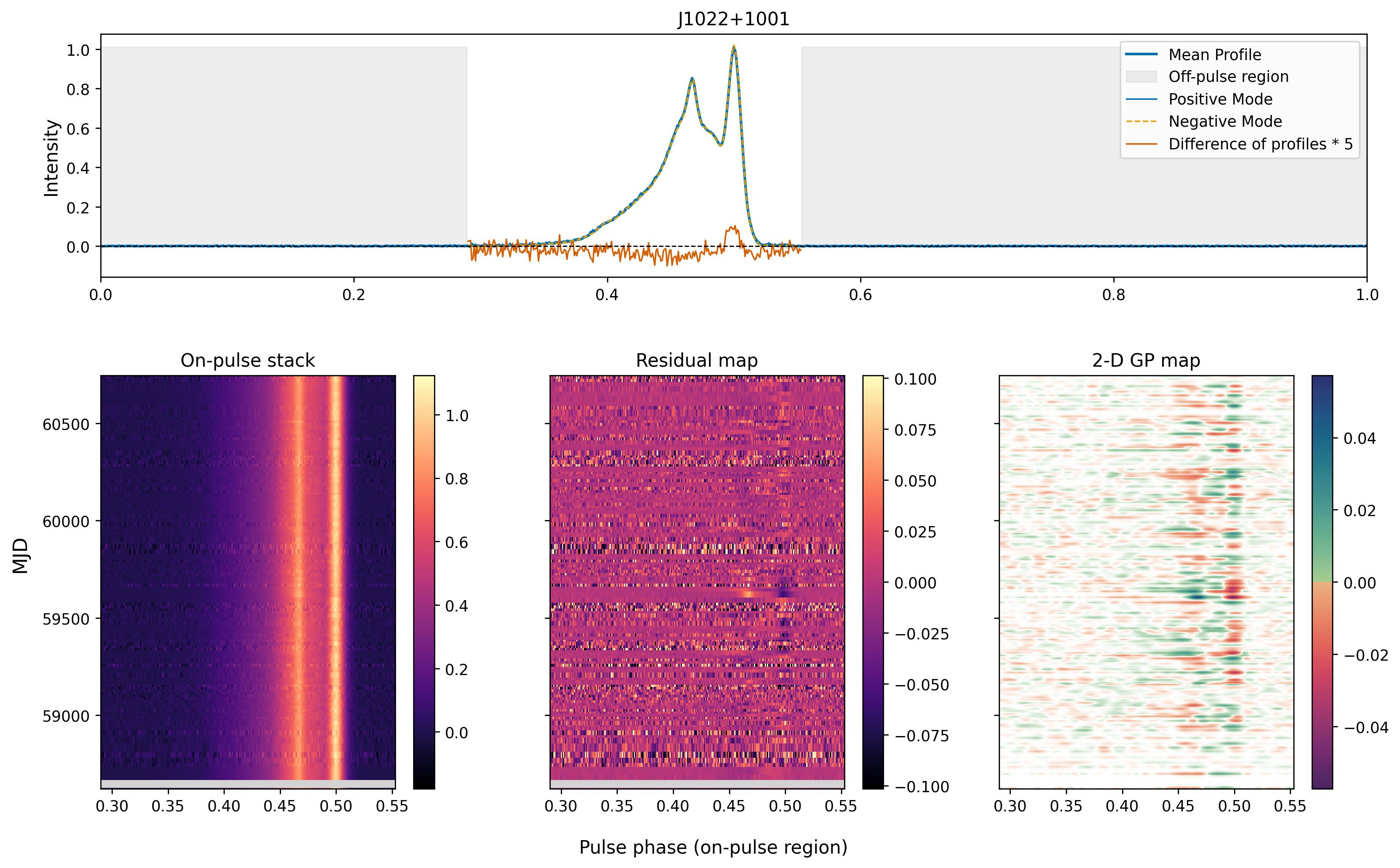}
    \caption{As Figure \ref{fig:J1713} but for PSR~J1022$+$1001. The variability is predominantly stochastic and is concentrated in the leading component of the pulse profile, consistent with the well-known profile instability of this pulsar.}
    \label{fig:J1022}
\end{figure*}

In our analysis, we observe the classic profile variations, but these appear broadly as epoch-to-epoch variations, with no clear long-term behaviour. The variability is primarily defined by the relative amplitude ratio of the leading and trailing peaks, consistent with previous studies (Figure~\ref{fig:J1022}). Given the relatively weak nature of the observed changes, it is possible that at least some of the variability is associated with pulse jitter, although we cannot conclusively distinguish between these possibilities with the present data.

\section{Discussion}
\label{sec:Discussions}
\subsection{Profile variation in the MPTA sample}
\label{sec:Behaviour}
The pulsars discussed in this section exhibit a diverse range of pulse-profile variability behaviours that can be broadly classified according to their temporal evolution. In several cases, the observed variability closely resembles phenomena previously reported in canonical pulsars~\citep{Lyne2010,Brook2016,Keith2025}. These include quasi-periodic profile variations, such as those observed in PSR~J2124$-$3358, isolated profile-change events followed by an approximately exponential recovery (e.g. PSRs~J1547$-$5709 and J1713$+$0747), and potentially repeating or recurrent profile change events, as seen in PSR~J0437$-$4715 (Mandow et al., in prep.). 
\begin{figure}
    \centering
    \includegraphics[width=\linewidth]{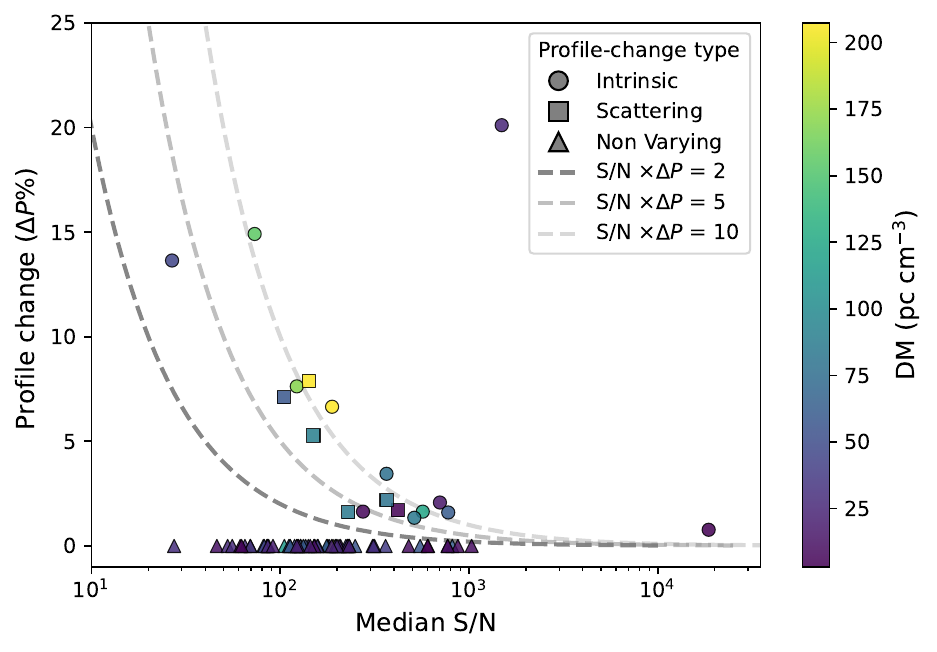}
    \caption{Measured pulse-profile change amplitude as a function of median observational S/N for the pulsars in our sample. Colours indicate the dispersion measure, while different marker styles distinguish intrinsic profile variability, scattering-induced variability, and non-varying pulsars. The dashed lines represent constant values of $\mathrm{S/N}\times \Delta P$, where $P$ is the fractional profile change. The distribution suggests an approximate detection threshold, with subtle profile variations requiring higher S/N for reliable identification by the current analysis pipeline.}
    \label{fig:SNR_ProfileChange}
\end{figure}

The observed distribution of pulse-profile variability is subject to a significant observational selection effect, as the minimum detectable profile change depends strongly on the signal-to-noise ratio (S/N) of each pulsar. Here, the S/N is defined as the median of the signal-to-noise ratios obtained with \texttt{pdmp} using \textsc{psrchive} from all individual observations of a given pulsar included in the analysis. The fractional profile-change amplitude, $\Delta P$, is defined as the maximum absolute difference between two representative pulse profiles, normalised by the peak intensity of the mean pulse profile.
For the pulsars with high S/N we are able to measure much smaller profile variations than those with lower S/N, implying that the observed sample of variable pulsars does not necessarily represent the intrinsic population. Figure~\ref{fig:SNR_ProfileChange} shows the measured fractional profile-change amplitude as a function of the median S/N of the MPTA observations. The distribution indicates that the transition from non-detection to detection occurs around $\mathrm{S/N}\times\Delta P\sim5$, although with considerable scatter. It also suggests that, for pulsars with median S/N $\lesssim300$, profile variations at the few-percent level remain largely below the sensitivity of the current analysis.

To infer the intrinsic distribution of profile variations while accounting for this selection effect, we model the profile-change amplitude, $\Delta P$, expressed as a percentage of the peak intensity of the mean pulse profile, as following a log-normal distribution parameterised by the logarithmic mean $\mu$ and standard deviation $\sigma$. Rather than adopting a hard detection threshold, we model the probability of detecting profile variations as a smooth function of both the median signal-to-noise ratio (S/N) of the observations and the profile-change amplitude,
\begin{equation}
P_{\rm det}(\Delta P,\mathrm{S/N})
=
\Phi\!\left(
\frac{\ln(\mathrm{S/N}\times\Delta P)-\ln C}{w}
\right),
\end{equation}
where $\Phi$ is the standard normal cumulative distribution function. Since $\Phi(0)=0.5$, the parameter $C$ (expressed in per cent) represents the characteristic value of $\mathrm{S/N}\times\Delta P$ at which the detection probability is 50 per cent, while the dimensionless parameter $w$ controls the width of the transition from non-detection to detection. The likelihood combines two types of information. For pulsars with measured profile-change amplitudes, it is proportional to the product of the probability density of the measured amplitude under the log-normal population model and the probability that a change of that amplitude would be detected at the pulsar's S/N. For non-detections, it includes the probability that the pulsar is either intrinsically stable or belongs to the variable population but remains below the detection sensitivity of the observations.

\begin{figure}
    \centering
    \includegraphics[width=\linewidth]{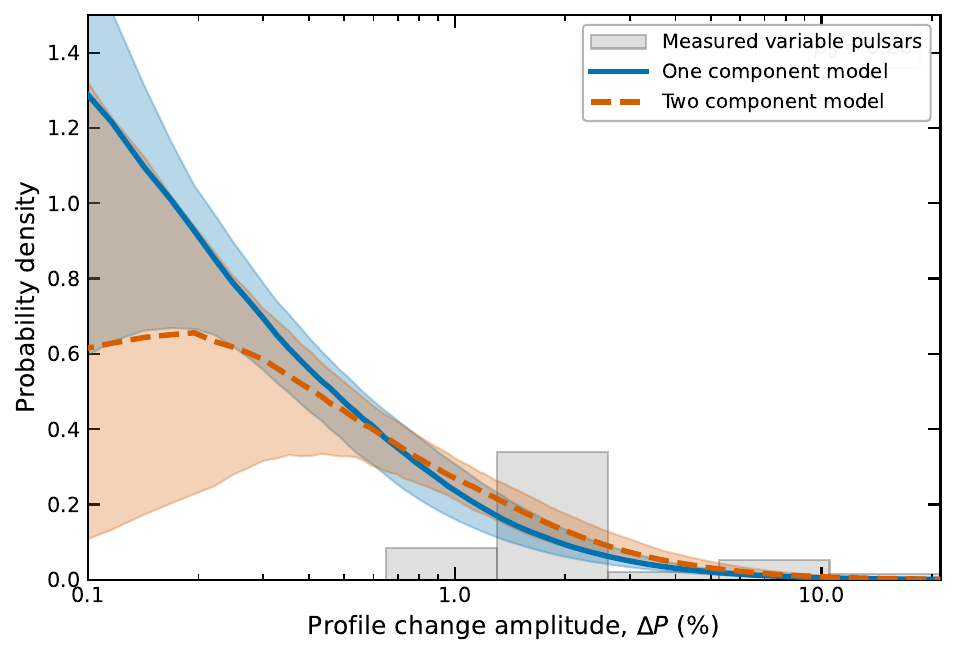}
    \caption{Inferred intrinsic distribution of pulse-profile variability amplitudes for millisecond pulsars. The dashed curve shows the posterior median of the single log-normal model, while the dotted curve shows the two-component population model, in which a fraction of pulsars exhibit no intrinsic profile variability and the remainder follow a log-normal distribution. The histogram shows the observed profile-change amplitudes for pulsars with detected variability.}
    \label{fig:Intrinsic_dist}
\end{figure}

We sample the posterior distribution using the \texttt{emcee} Markov Chain Monte Carlo sampler \citep{EMCEE2013}, obtaining $\mu = -0.5^{+0.5}_{-0.6}$, $\sigma = 1.6^{+0.4}_{-0.3}$, $C = 5.1^{+2.6}_{-1.3}$, and $w = 0.5^{+0.4}_{-0.2}$. The inferred intrinsic distribution, shown in Figure~\ref{fig:Intrinsic_dist}, indicates that low-level profile variations dominate the population, with approximately 48 per cent of intrinsically variable MSPs exhibiting fractional profile changes of 1 per cent or less. Detecting variability at this level typically requires observations with S/N exceeding $\sim500$, explaining why many such pulsars remain undetected in the current sample.

To constrain the overall fraction of MSPs that exhibit intrinsic profile variability, we additionally consider a two-component population model in which a fraction $f_{\rm z}$ of pulsars have intrinsically stable profiles, while the remaining fraction $(1-f_{\rm z})$ belongs to the variable population described by the log-normal distribution. 
The complete likelihood for this model is 
\begin{equation}
\begin{split}
\mathcal{L}
&={}
\prod_{i\in\mathcal{D}}
(1-f_{\rm z})\,
p(\Delta P_i\mid\mu,\sigma)\,
P_{\rm det}(\Delta P_i,S_i)
\\
&\times
\prod_{j\in\mathcal{N}}
\left[
f_{\rm z}
+
(1-f_{\rm z})
\int_{0}^{\infty} \! p(x\mid\mu,\sigma)
\left[1-P_{\rm det}(x,S_j)\right]
\,{\rm d}x
\right],
\end{split}
\label{eq:population_likelihood}
\end{equation}
where $\mathcal{D}$ and $\mathcal{N}$ denote the sets of pulsars with detected and non-detected profile variations, respectively, and $S_i$ denotes the median S/N of pulsar $i$. The first term accounts for detected pulsars, where the likelihood includes the probability $(1-f_{\rm z})$ that it belongs to the variable population, the probability density $p(\Delta P_i\mid\mu,\sigma)$ of its measured profile-change amplitude under the intrinsic log-normal distribution, and the probability $P_{\rm det}(\Delta P_i,S_i)$ of detecting that variation. For a non-detection, the second term accounts for two possibilities: the pulsar may have zero intrinsic variability, with probability $f_{\rm z}$, or it may belong to the variable population but have a profile-change that remains undetected. The latter probability is obtained by marginalising $1-P_{\rm det}$ over the intrinsic log-normal distribution of $\Delta P$. The posterior distribution in this case implies that at least 40 per cent of MSPs exhibit intrinsic pulse-profile variability at the 95 per cent credible level. Whilst the intrinsic distribution (Figure~\ref{fig:Intrinsic_dist}) shifts slightly, the overall effect is the same.

\subsection{Scattering induced profile variations}
\label{sec:scattering_var_pulsar}
High-DM pulsars with sufficiently high S/N were particularly sensitive targets for identifying scattering-induced profile evolution, particularly when the MPTA noise analysis favoured timing models including chromatic noise components~\citep{MPTA_data}. Such chromatic propagation effects can mimic intrinsic pulse-profile changes, often appearing as smooth, slowly evolving structures in the residual maps.
To distinguish intrinsic profile changes from frequency-dependent propagation effects, we applied two diagnostic tests to every pulsar exhibiting significant profile variability.

First, we examined the frequency dependence of the profile variations using the full bandwidth of the MeerKAT L-band observations. The observations were divided into frequency subbands, and the profile evolution was inspected across the band. Across the MeerKAT L-band, intrinsic profile changes may show weaker frequency dependence than scattering-induced broadening, although intrinsic profile evolution can itself be frequency dependent, whereas scattering-induced variations should become increasingly pronounced towards lower observing frequencies. For each pulsar, we quantified the frequency dependence by measuring the amplitude of the profile variation as a function of observing frequency. Pulsars exhibiting a steep frequency dependence were classified as scattering candidates.

\begin{figure}
    \centering
    \includegraphics[width=\linewidth]{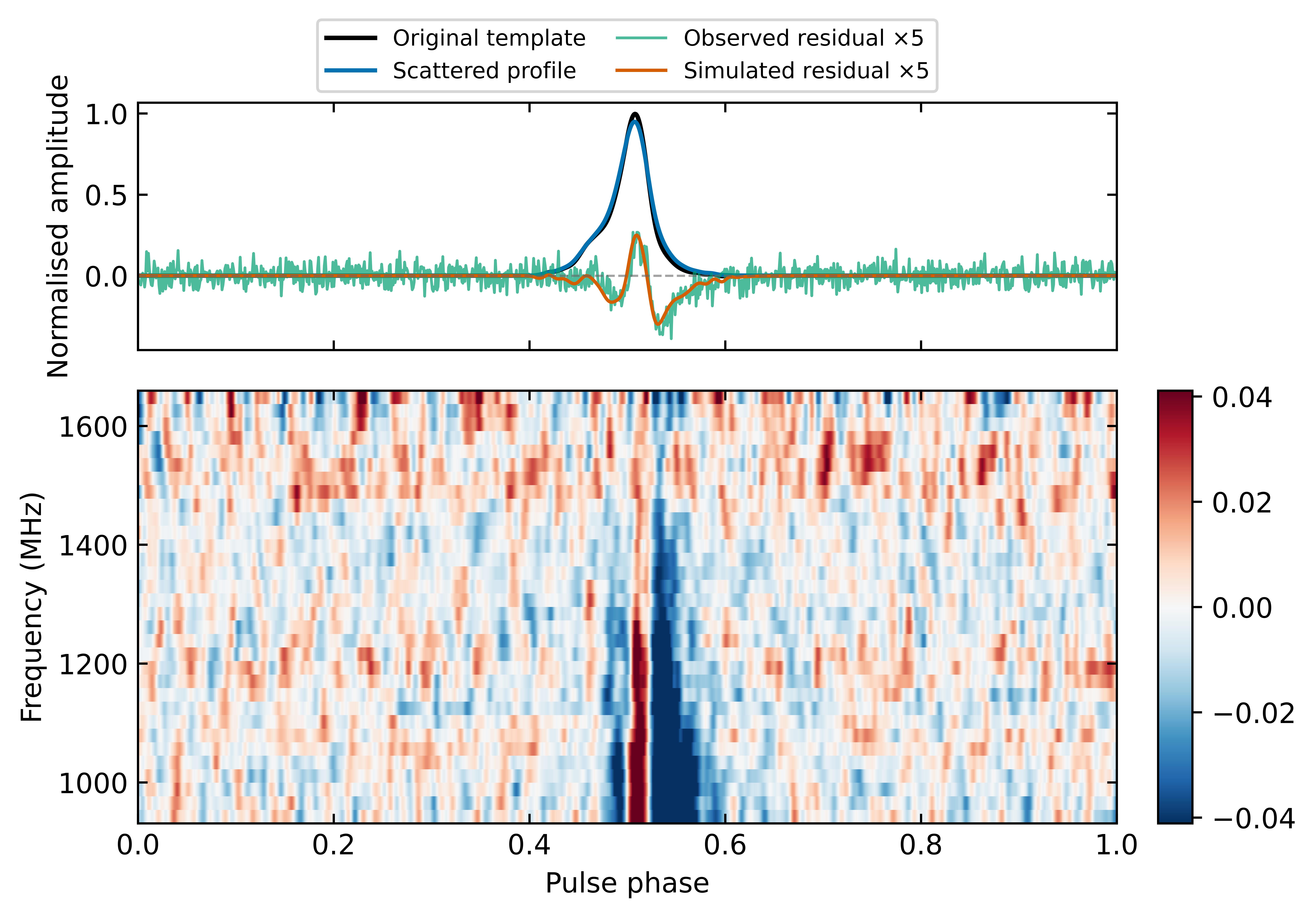}
    \caption{Scattering analysis for PSR~J1431$-$5740. \textit{Top panel:} Comparison between the original pulse-profile template (black) and the profile obtained after convolving the frequency-resolved portrait with an exponential scattering kernel (blue). The observed profile residual (green) and the simulated scattering residual (orange), both scaled by a factor of five for clarity, show similar phase-dependent morphology, indicating that the observed profile variation is consistent with scattering-induced broadening. \textit{Bottom panel:} Frequency dependence of the observed profile residual across the L-band after subtraction of the average profile. The variability increases towards lower observing frequencies, as expected for interstellar scattering, supporting a chromatic origin of the observed pulse-profile evolution.}
    \label{fig:J1431_frequency}
\end{figure}

Second, for these candidates we modelled the expected effects of interstellar scattering by convolving the frequency-resolved pulse portrait with a one-sided exponential scattering kernel. The resulting scattered profiles were compared with the observed profile variations through their residuals relative to the original template. A close agreement between the simulated and observed residual morphology, together with a strong frequency dependence, was taken as evidence that the observed variability is dominated by interstellar scattering rather than intrinsic emission changes.
We found that for these pulsars the simulated profile residual closely reproduced the phase-dependent morphology of the observed profile residuals, as shown in the upper panel of Figure~\ref{fig:J1431_frequency}. This agreement suggests that the dominant source of variability in the pulsars discussed in Appendix~\ref{sec:Scattering_cases} is most likely due to frequency-dependent interstellar propagation effects.

We further examined the distribution of profile variability across the full pulsar sample as a function of dispersion measure and median signal-to-noise ratio. As shown in Figure~\ref{fig:SNR_ProfileChange}, pulsars classified as scattering candidates tend to preferentially occupy the high-DM, high-S/N region of parameter space, supporting the interpretation that scattering-induced profile variability is more readily detectable in such systems. Nevertheless, not all high-DM pulsars exhibit measurable variability, suggesting that additional factors, such as line-of-sight turbulence or intrinsic pulse morphology, also influence the detectability of scattering signatures.

\subsection{Profile changes and timing noise}
Several studies of canonical pulsars have reported correlations between variations in the spin-down rate, $\dot{\nu}$, and changes in the average pulse profile~\citep{Lyne2010,Brook2016,Shaw2022,Basu2024,Lower2025,Keith2025}. 
Such variations in $\dot{\nu}$ are thought to contribute to the observed \textit{achromatic timing noise} in pulsars~\citep{Lyne2010,Brook2016}, typically manifesting as low-frequency, quasi-quadratic structures in the timing residuals. 

Measuring correlations between pulse-profile variability and changes in the spin-down rate, $\dot{\nu}$, is particularly challenging in millisecond pulsars, since their spin-down rates are typically very small, of the order of $\lesssim 10^{-16}~\mathrm{Hz\,s^{-1}}$, and it has been observed that the $\dot{\nu}$ fluctuations seem to scale directly with $\dot{\nu}$, at a level of around 0.1--1 per cent \citep{Lower2025}.
Whilst the timing precision for MSPs should help, the noise budget for the best timed PTA pulsars is dominated by propagation effects from the interstellar medium, such as dispersion and scattering.
Furthermore, any profile changes themselves directly add additional time-correlated noise that can further mask the ability to search for $\dot{\nu}$ variations.
In contrast, the canonical pulsars that have been studied are typically dominated by the timing noise, making it much easier to detect the underlying $\dot{\nu}$ variations.

We attempted to measure the $\dot{\nu}$ variations for the three pulsars with the clearest quasi-periodic variations, PSRs~J0900$-$3144, J1103$-$5403 and J2124$-$3358.
In all cases the $\delta\dot{\nu}$ is consistent with zero, with uncertainties of around $10^{-18}$, larger than the expected $\delta\dot{\nu}$ based on \citet{Lower2025}.
To investigate this further, we progressively increased the length of the timing dataset in steps of 0.5\,yr and performed a complete timing analysis for each subset. For each dataset, we fitted the timing model together with the stochastic noise components adopted in the MPTA analysis~\citep{MPTA_data}, including achromatic red noise, dispersion-measure noise, and, where required, chromatic noise. The resulting best-fit timing model was then used as input to a weighted least-squares fit with \textsc{TEMPO2}~\citep{TEMPO2A}, from which we measured the uncertainty on the recovered spin-down rate, $\dot{\nu}$, as a function of the observing span.

\begin{figure}
    \centering
    \includegraphics[width=\linewidth]{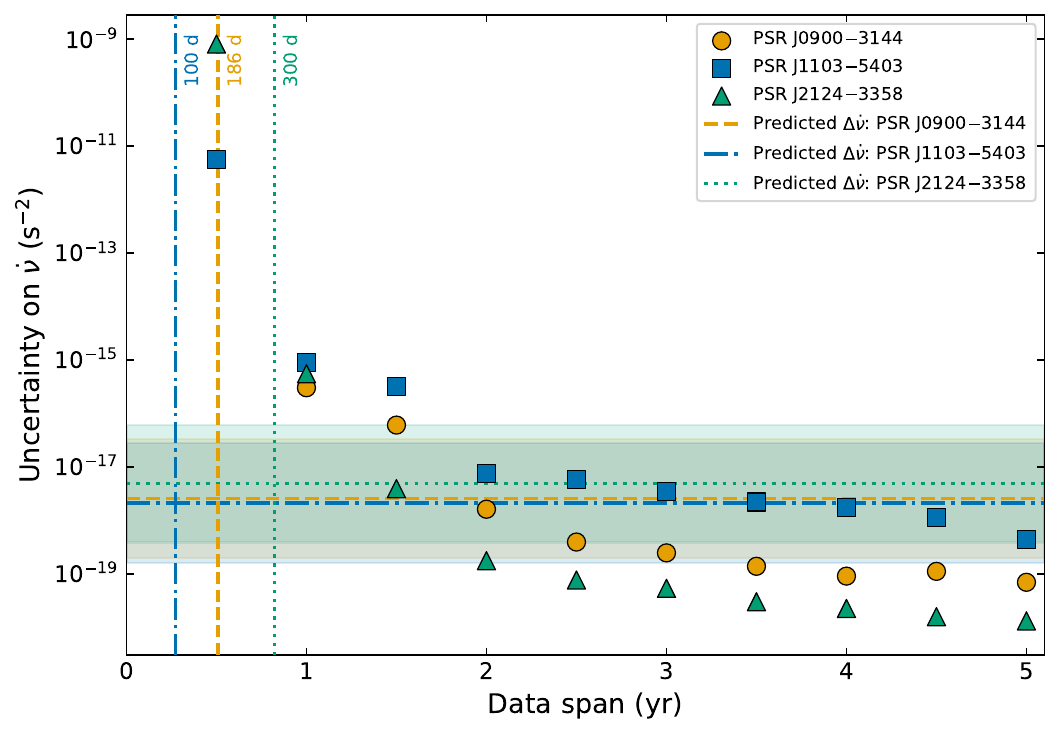}
        \caption{Uncertainty on the measured spin-down rate, $\dot{\nu}$, as a function of data span for PSRs~J0900$-$3144 (orange circles), J1103$-$5403 (blue squares), and J2124$-$3358 (green triangles). The horizontal lines indicate the expected magnitude of the spin-down-rate variation, $\Delta\dot{\nu}$, predicted from the empirical relation of \citet{Lower2025}, with different colours and line styles corresponding to the three pulsars. The shaded regions denote the uncertainty range obtained by propagating the uncertainties in the empirical relation. Vertical lines mark the search-window durations adopted for each pulsar (100, 186, and 300\,d, respectively) from the visual inspection of the profile change. As the data span increases, the uncertainty on $\dot{\nu}$ decreases and approaches, or falls below, the expected amplitude of the spin-down-rate variation, illustrating the observing baseline required to detect $\Delta\dot{\nu}$ at the predicted level.}
    \label{fig:error_on_nudot_measurement}
\end{figure}

In Figure~\ref{fig:error_on_nudot_measurement}, we show the expected uncertainty in $\dot{\nu}$ as a function of timing span for three representative pulsars showing quasi periodic behaviour in our sample: PSRs~J1103$-$5403, J0900$-$3144, and J2124$-$3358. Also shown are the values of $\delta\dot{\nu}$ predicted by the empirical relation of \citet{Lower2025}. 
The uncertainty decreases rapidly as the timing baseline increases, indicating that a statistically significant measurement of the expected spin-down-rate variation would typically require the pulsar to remain in the same emission state for approximately $1$--$3$\,yr, depending on its timing precision and noise properties. 
 We find that this estimate is consistent with the uncertainties obtained independently from the Gaussian-process\footnote{\href{https://gitlab.com/benjaminshaw/pulsarpvc}{https://gitlab.com/benjaminshaw/pulsarpvc}}~\citep{Shaw2022} analysis of the timing residuals, which yields $\dot{\nu}$ uncertainties of order $10^{-19}\,\mathrm{Hz\,s^{-1}}$ over comparable timescales.
Since the profile-state transitions identified in this work occur on substantially shorter timescales, the associated changes in $\dot{\nu}$ are expected to remain unresolved with current datasets and analysis techniques. 
Future observations with longer timing baselines, together with analysis methods that explicitly incorporate the observed profile-change epochs when fitting for changes in $\dot{\nu}$, will improve the sensitivity to such measurements. These developments will be essential for determining whether the profile variations observed in millisecond pulsars are accompanied by changes in spin-down rate, as is observed in canonical pulsars.

We also performed a similar analysis on one of the most significant profile-variable pulsars, PSR~J1547$-$5709 (Figure~\ref{fig:J1547}). By fitting only for a change in spin-down rate together with DM noise, we obtained an estimate of the spin-down-rate variation of $\delta\dot{\nu} = -8.8 \pm 3.4\times10^{-19}$, with the corresponding transition epoch constrained to MJD~59571(191). While not conclusive, this result suggests the possibility of an underlying change in $\dot{\nu}$ associated with the observed pulse-profile variation in this pulsar.

\subsection{Profile-associated \texorpdfstring{$\Delta \dot \nu$}{} variations as a potential source of achromatic timing noise}
Although we are unable to directly detect $\delta\dot{\nu}$ in our sample, several pulsars exhibit clear quasi-periodic profile variations. Here we estimate the level of achromatic timing noise that could arise if these profile changes are associated with variations in the spin-down rate ($\dot{\nu}$).
To quantify the profile variability, we performed a principal component analysis (PCA) on the two-dimensional Gaussian Process reconstruction of the profile-variability maps of PSR J2124$-$3358 following the process in \citet{Keith2025}.
The first principal component, represented by its corresponding temporal coefficients, captures the majority of the observed profile variability and therefore provides a convenient proxy for the evolution of the profile state.

We used the temporal coefficients of the first principal component as a template for simulated spin-down-rate variations, scaling the amplitude to prescribed values of $\delta\dot{\nu}$.
Simulated timing datasets were generated with the \textsc{toasim} framework distributed with \textsc{tempo2} \citep{TEMPO2A}, including only white noise, dispersion measure variations and the injected $\delta\dot{\nu}$.
We generated two sets of simulations, with $\delta\dot{\nu}=10^{-18}\,\mathrm{s}^{-2}$ and $10^{-19}\,\mathrm{s}^{-2}$, corresponding to approximately 1 per cent and 0.1 per cent of the pulsar's secular spin-down rate. These values span the range predicted by \citet{Lower2025}.

\begin{figure}
    \centering
    \includegraphics[width=\linewidth]{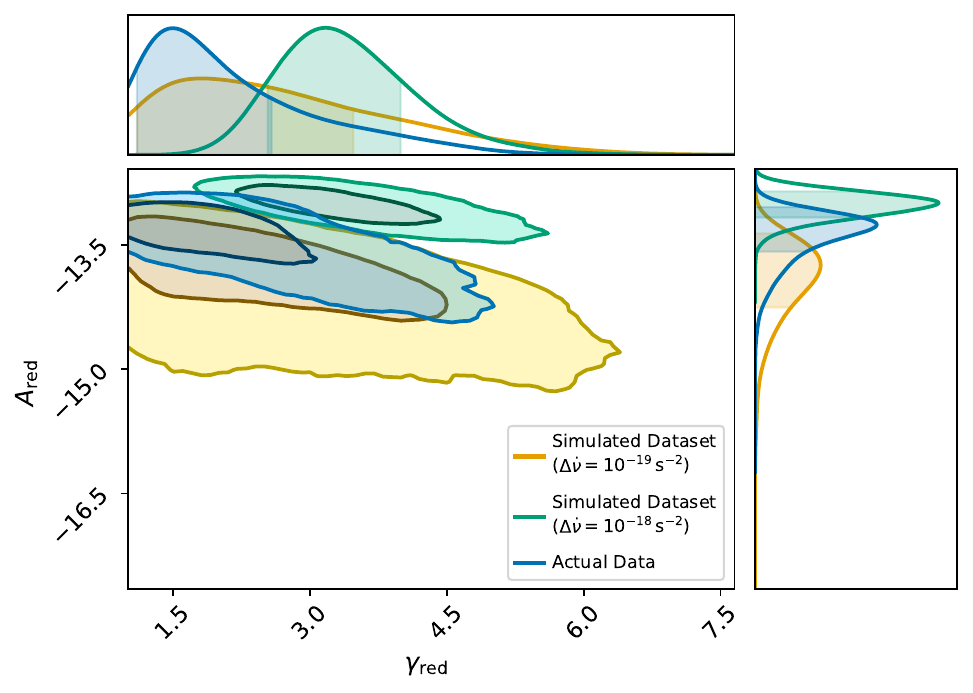}
    \caption{Comparison of the red-noise amplitude ($A_{\mathrm{red}}$) and spectral index ($\gamma$) recovered from the actual timing data of PSR J2124$-$3358 and from simulated datasets with injected spin-down-rate variations of $\delta\dot{\nu}=10^{-19}$~s$^{-2}$ and $10^{-18}$~s$^{-2}$. The comparison illustrates how different levels of injected $\delta\dot{\nu}$ can produce red-noise characteristics that can overlap those inferred from the observed data, providing insight into whether unresolved spin-down-rate switching can account for the measured achromatic red noise.}
    \label{fig:nudot_variation_sim}
\end{figure}

We then analysed the simulated data using \texttt{enterprise} to fit for the standard Fourier basis red noise commonly used in PTA analysis \citep{Lentati2013,EnterpriseOrg}.
The model is parametrised by an amplitude $A_\mathrm{red}$ and spectral exponent $\gamma$.
We solved for the posterior distributions of the model hyperparameters using \texttt{dynesty}~\citep{DynestySampelr} via \texttt{run\_enterprise} \citep{Keith2023}. The resulting posterior distributions are shown in Figure~\ref{fig:nudot_variation_sim}, alongside the posterior obtained from the actual data from our six-year dataset.

For $\delta\dot{\nu}=10^{-18}$~s$^{-2}$, the inferred spectral exponent is similar to that measured for MSPs exhibiting significant timing noise. However, the corresponding red-noise amplitude is substantially larger than that measured for PSR J2124$-$3358 and lies near the upper end of the PTA pulsar population, where typical values are $\log A_\mathrm{red}\sim -14$ and $\gamma\approx3$ - 4 (e.g.\ \citealp{MPTA_data,EPTA_noise_ppr}).

For $\delta\dot{\nu}=10^{-19}\,\mathrm{s}^{-2}$, the inferred red-noise amplitude decreases accordingly, consistent with the expectation that the amplitude of the induced timing noise scales with the magnitude of the injected spin-down variations. The posterior favours flatter values of $\gamma$, which is commonly observed for low-significance red-noise detections in PTA analyses. 
Notably, the posterior recovered from the $\delta\dot{\nu}=10^{-19}\,\mathrm{s}^{-2}$ simulation broadly overlaps with that inferred from the actual timing data of PSR J2124$-$3358 in both $A_{\rm red}$ and $\gamma$ (Figure~\ref{fig:nudot_variation_sim}). This suggests that spin-down-rate variations at this level can produce red-noise properties consistent with those observed for PSR J2124$-$3358.
This flattening is therefore likely a consequence of limited signal-to-noise ratio rather than a physical correlation between $\gamma$ and $\delta\dot{\nu}$. With improved sensitivity to the red-noise process, the inferred spectral exponent is expected to converge towards steeper values, with the corresponding amplitude approaching the region occupied by many PTA pulsars ($\gamma\sim3$ - 4 and $\log A_\mathrm{red}\sim-14$). These simulations demonstrate that low-level quasi-periodic $\dot{\nu}$ variability (of order 0.1 per cent) could plausibly contribute to the weak achromatic timing noise observed in PTA datasets whilst remaining hard to detect directly.

\section{Conclusions}
We have carried out a  systematic search for long-term pulse-profile variability in the MeerKAT Pulsar Timing Array, analysing six-years of observations of 84 millisecond pulsars using a two-dimensional Gaussian-process framework. 
This approach enables coherent, low-amplitude profile evolution to be separated from stochastic observational noise and reveals a broader range of long-term time-correlated behaviour than has previously been recognised in MSPs.

We identify pulse-profile variability in 18 pulsars, spanning several phenomenological categories, including quasi-periodic modulation, isolated profile-change events, some of which are followed by approximately exponential recovery, stochastic profile evolution, and long-term secular changes. 
Six pulsars show strong frequency-dependent variability whose morphology can be reproduced by scattering simulations, indicating that their profile evolution is most likely dominated by chromatic propagation effects in the interstellar medium. 
The remaining 12 pulsars exhibit predominantly frequency-independent variability, favouring an intrinsic origin associated with changes in the pulsar emission process. 

The detectability of these changes depends strongly on the signal-to-noise ratio of the observations. After accounting for this selection effect, we infer that at least 40 per cent of MSPs exhibit intrinsic pulse-profile variability at the 95 per cent credible level. The inferred variable population is dominated by low-amplitude changes, with approximately 48 per cent of intrinsically variable MSPs expected to exhibit fractional profile variations of 1 per cent or less. Many such variations therefore remain below the detection threshold of the present MPTA dataset, implying that the observed sample represents only the most readily detectable part of the underlying variable population.

Motivated by the established association between profile-state changes and spin-down-rate variations in canonical pulsars, we searched for corresponding changes in $\dot{\nu}$ in the most variable MSPs. 
We find no statistically significant measurements of $\delta\dot{\nu}$, with the recovered values remaining consistent with zero within uncertainties of order $10^{-18}\,\mathrm{s}^{-2}$. 
Our sensitivity analysis indicates that detecting variations at the level expected from empirical scaling relations would generally require individual emission states to persist for approximately 1--3 yr, depending on the timing precision and noise properties of the pulsar. 
This is longer than the characteristic state durations observed for several of the quasi-periodic sources in our sample.

Although the expected spin-down-rate variations cannot currently be detected directly, our simulations show that unresolved, low-amplitude variations in $\dot{\nu}$ could plausibly contribute to a fraction of the weak low-frequency timing noise observed in some PTA pulsars.
In particular, spin-down-rate variations with amplitudes of order $10^{-19}$--$10^{-18}\,\mathrm{s}^{-2}$ could plausibly contribute to the weak low-frequency timing noise while remaining difficult to distinguish from other stochastic processes. 
This result does not establish a common physical origin for pulse-profile variability and achromatic timing noise, but it demonstrates that such a connection is consistent with the present data.

The principal conclusion of this work is that long-term time-correlated pulse-profile variability is not restricted to a small number of MSPs, but may be a common phenomenon whose observed occurrence rate is presently limited by sensitivity. 
As longer MeerKAT timing baselines become available and future telescope deliver substantially higher signal-to-noise observations, increasingly subtle profile variations will become detectable. 
Robust modelling of both intrinsic pulse-shape evolution and chromatic propagation effects will therefore be essential for limiting systematic timing errors and realising the full sensitivity of next-generation pulsar timing arrays to nanohertz gravitational waves.

\section*{Acknowledgements}
The author acknowledge Avishek Basu for exciting discussion on variability in canonical pulsars. The author also acknowledge Rami Mandow for discussion and feedback regarding the draft. The author acknowledges the referee for their insightful comments.
BB acknowledge funding from the United Kingdom’s Research and Innovation (UKRI) Science and Technology Facilities Council (STFC), project reference [ST/Y509814/1]. This work was partially supported by the Australian Research Council Centre of Excellence for Gravitational Wave Discovery (CE230100016).
The MeerKAT telescope is operated by the South African Radio Astronomy Observatory, which is a facility of the National Research Foundation, an agency of the Department of Science and Innovation.
\section*{Data Availability}
All calibrated and RFI-cleaned observations used in this study can be downloaded from the Pulsar Portal (\url{https://pulsars.org.au}) after the 1.5-year embargo period following the date of observation (Bailes et. al 2026 (in prep)).
 



\bibliographystyle{mnras}
\bibliography{reference} 




\appendix
\section{Pulsars showing profile shape changes likely due to scattering variations}
These pulsars exhibit pulse-profile variations that evolve significantly with observing frequency across the L-band, producing a steep frequency dependence in the measured profile changes. Such behaviour may arise from strong chromatic propagation effects, particularly interstellar scattering, although intrinsic changes in the pulsar emission cannot be completely ruled out. A more detailed discussion of these sources is presented in Section~\ref{sec:scattering_var_pulsar}. In this section, we present the pulsars whose observed variability is more consistent with scattering-induced effects than with intrinsic pulse-profile evolution.
\label{sec:Scattering_cases}
\subsection{PSR J1431$-$5740}
PSR~J1431$-$5740 is a high-dispersion measure millisecond pulsar with DM$\sim$131~pc~cm$^{-3}$. Recent work by~\citet{Kulkarni2025} has suggested the possible presence of a scattering-related event in this pulsar, based on comparisons between chromatic timing processes and flux-density variations.

\begin{figure*}
    \centering
    \includegraphics[width=0.86\linewidth]{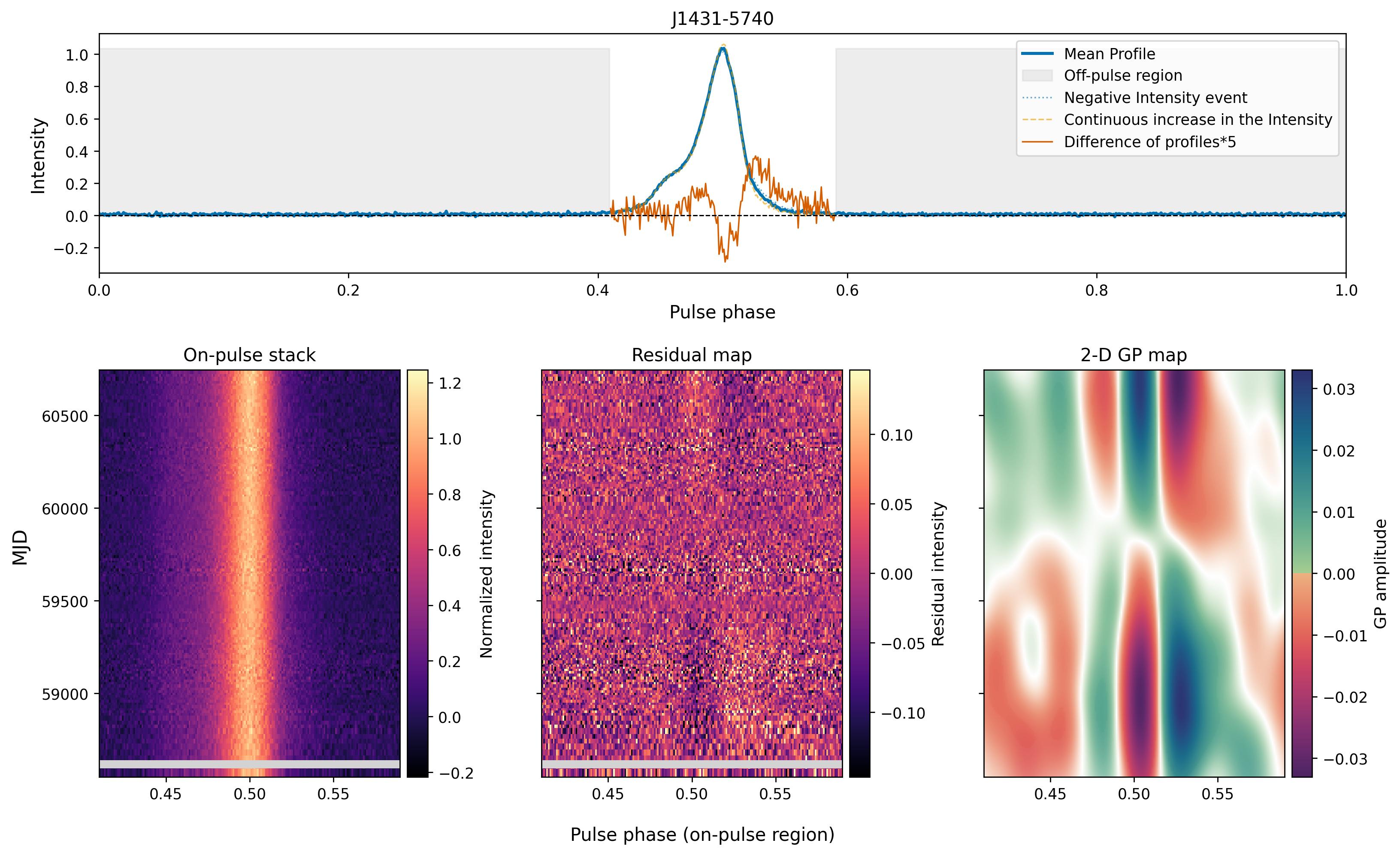}
    \caption{As Figure \ref{fig:J1713} but for PSR~J1431$-$5740. The profile evolution is characterised by smooth, long-timescale intensity variations concentrated around the peak and trailing edge of the pulse profile.}
    \label{fig:J1431}
\end{figure*}

In our analysis, the pulsar exhibits a gradual and smooth evolution of the pulse profile over long timescales, primarily concentrated around the peak and trailing edge of the profile, as shown in Figure~\ref{fig:J1431}. The observed variability corresponds to an intensity change of approximately $\sim 2.5$ per cent. Unlike abrupt mode-changing behaviour, the evolution appears continuous and correlated in time.

\subsection{PSR J1017$-$7156}
\begin{figure*} 
    \centering
    \includegraphics[width=0.86\linewidth]{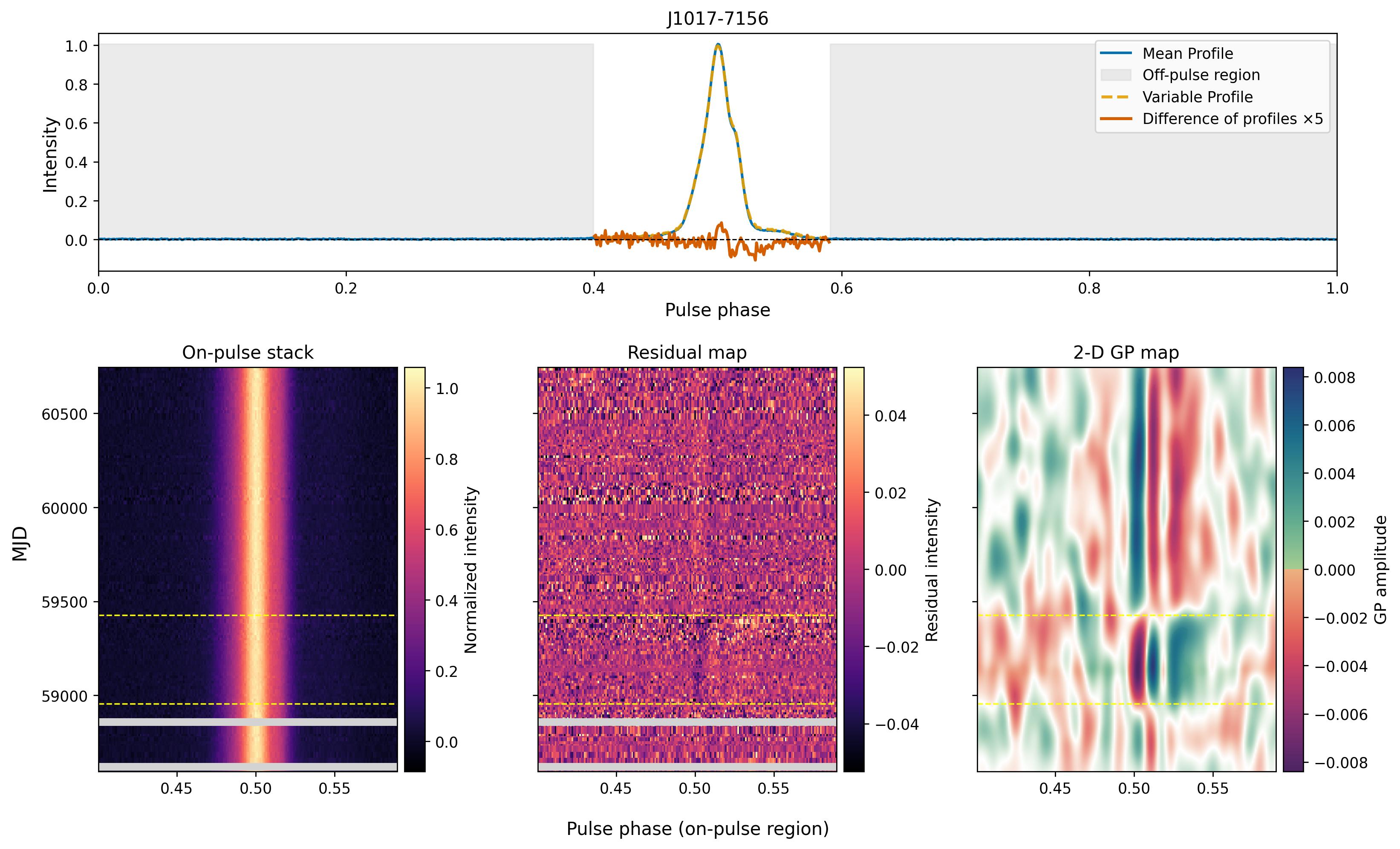}
    \caption{As Figure \ref{fig:J1713} but for PSR~J1017$-$7156. A localized one-off profile-change event is observed between the two dashed horizontal lines, with the variability confined primarily to the peak and trailing edge of the pulse profile. This event could be associated with an Extreme Scattering Event (ESE). }
    \label{fig:J1017}
\end{figure*}

PSR~J1017$-$7156 has shown strong variation in the past, owing to a turbulent interstellar structure along the line of sight~\citep{Coles2015}. This pulsar exhibits a profile-change event between MJD~58881.1 and 59431.4, which can be characterised as a one-off transition in the pulse profile. The observed variation is subtle, followed by a gradual recovery towards the nominal profile shape. 
While only faint, temporally correlated structure is visible in the residual map (Figure~\ref{fig:J1017}), the two-dimensional Gaussian process reconstruction clearly captures this variation.

\subsection{PSR J1708$-$3506}
\begin{figure*}
    \centering
    \includegraphics[width=0.86\linewidth]{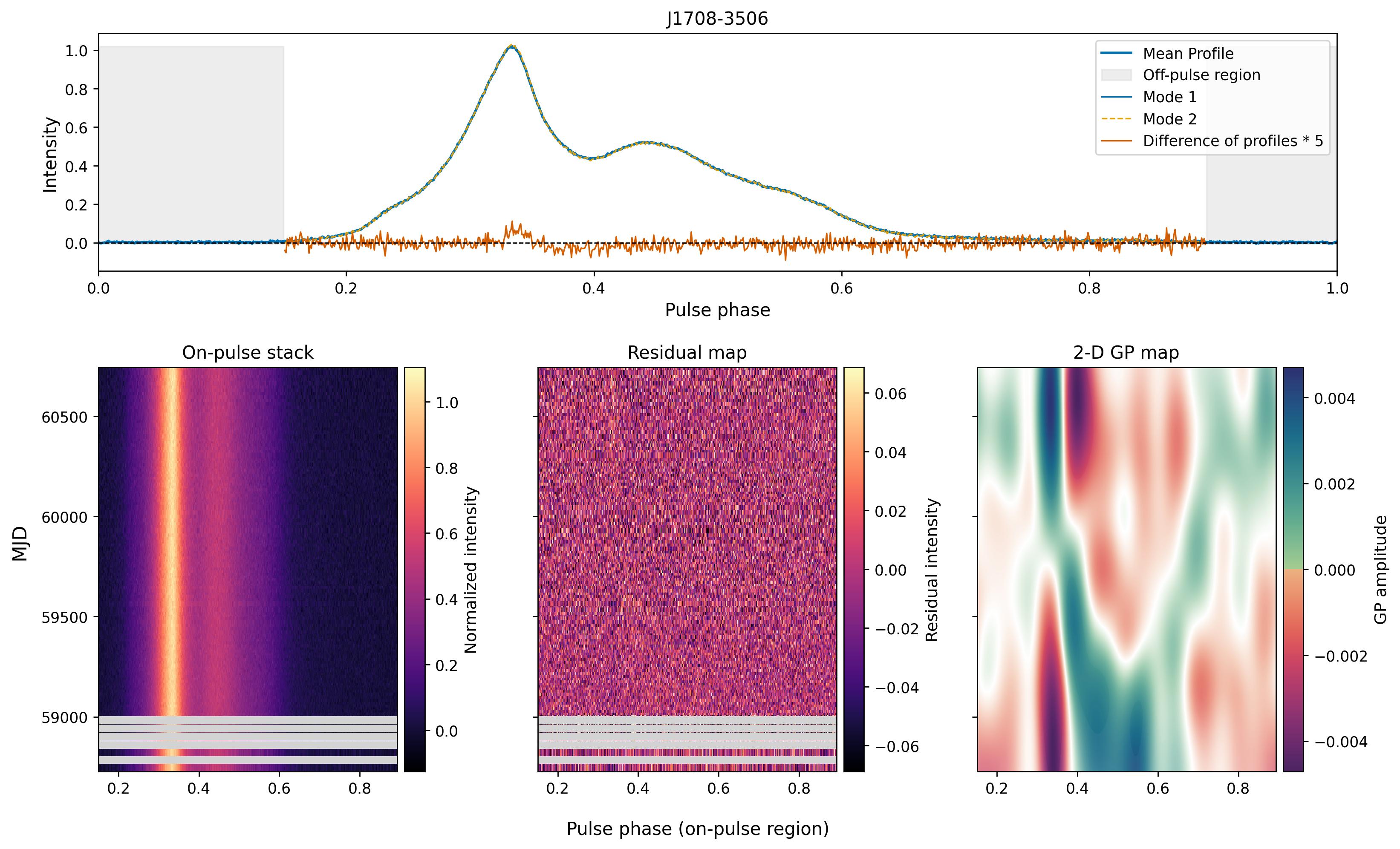}
    \caption{As Figure \ref{fig:J1713} but for PSR~J1708$-$3506. The pulsar exhibits weak, smoothly varying profile evolution distributed over the full observing span, producing coherent long-timescale changes across the main pulse component.}
    \label{fig:J1708}
\end{figure*}

In our analysis, we detect a subtle but statistically significant long-term evolution in the pulse profile. 
Unlike abrupt mode-changing or transient events seen in some other pulsars, the variation in PSR~J1708$-$3506 appears as a gradual secular change across the observing span. 
The profile slowly evolves from a relatively higher-intensity state at the beginning of the dataset to a lower-intensity state towards the end of the dataset, as shown in Figure~\ref{fig:J1708}. 
The variation is primarily localised around the leading peak and the trailing shoulder of the profile, with an overall intensity change of only a few percent.  

\subsection{PSR J1747$-$4036}

In our analysis, PSR~J1747$-$4036 exhibits slow secular pulse-profile evolution rather than abrupt mode switching or short-lived transient events. 
The variability is dominated by gradual changes in the leading component of the profile, which shows a monotonic decrease in intensity over the observing span.
In addition, weaker correlated changes are visible across the broader trailing component. The evolution appears smooth and continuous, resembling a gradual transition between higher- and lower-intensity states over long timescales.

\begin{figure*}
    \centering
    \includegraphics[width=0.86\linewidth]{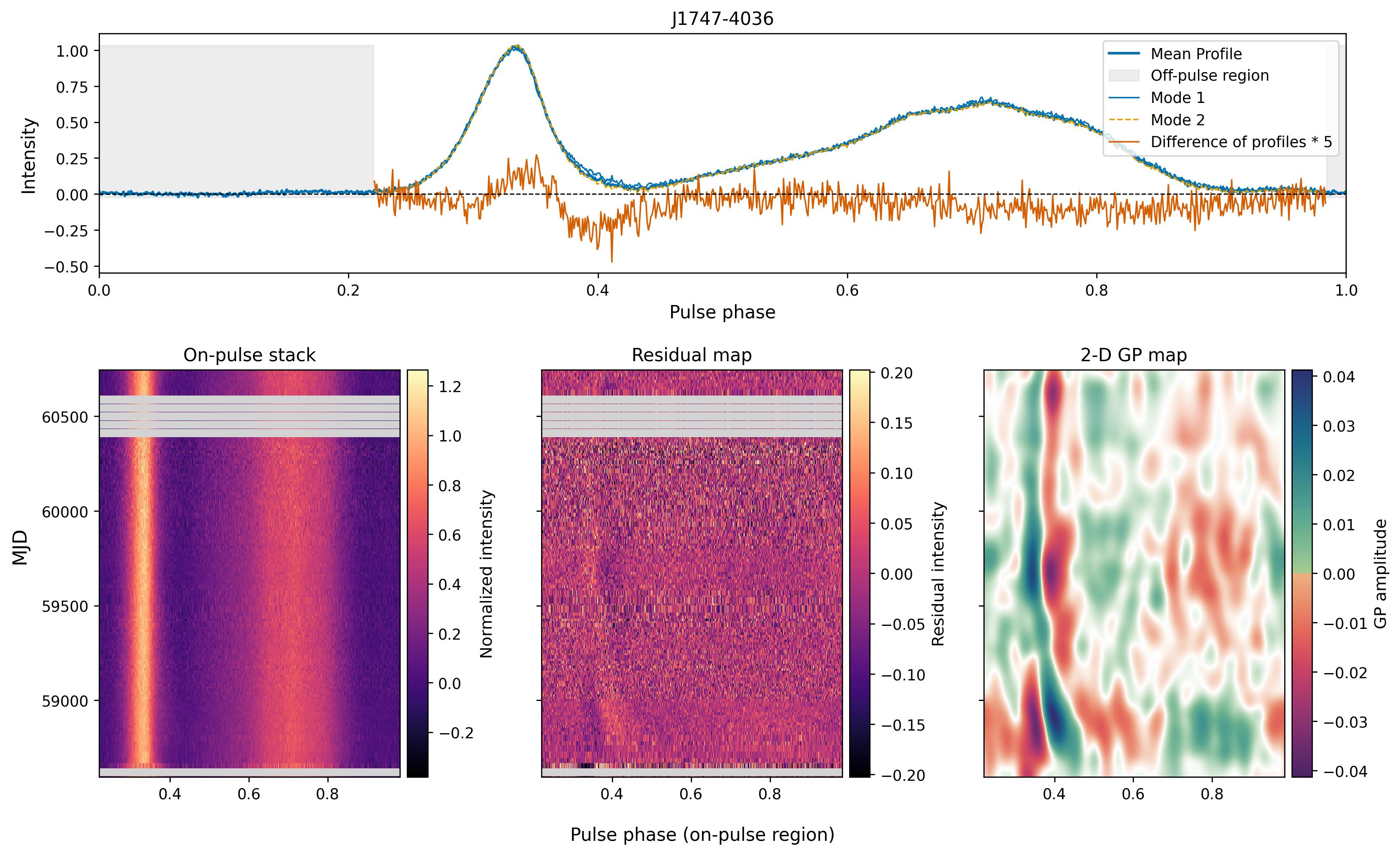}
    \caption{As Figure \ref{fig:J1713} but for PSR~J1747$-$4036. The variability is dominated by slow, coherent evolution across the broad trailing component, producing a gradual redistribution of pulse intensity over the observing span.}
    \label{fig:J1747}
\end{figure*}

This behaviour is also evident in the residual map shown in Figure~\ref{fig:J1747}, where correlated structures persist across multiple observing epochs. The two-dimensional Gaussian-process reconstruction further highlights coherent long-timescale variability, suggesting that the observed profile evolution is not dominated by white noise. 

\subsection{PSR J1802$-$2124}

\begin{figure*}
    \centering
    \includegraphics[width=0.86\linewidth]{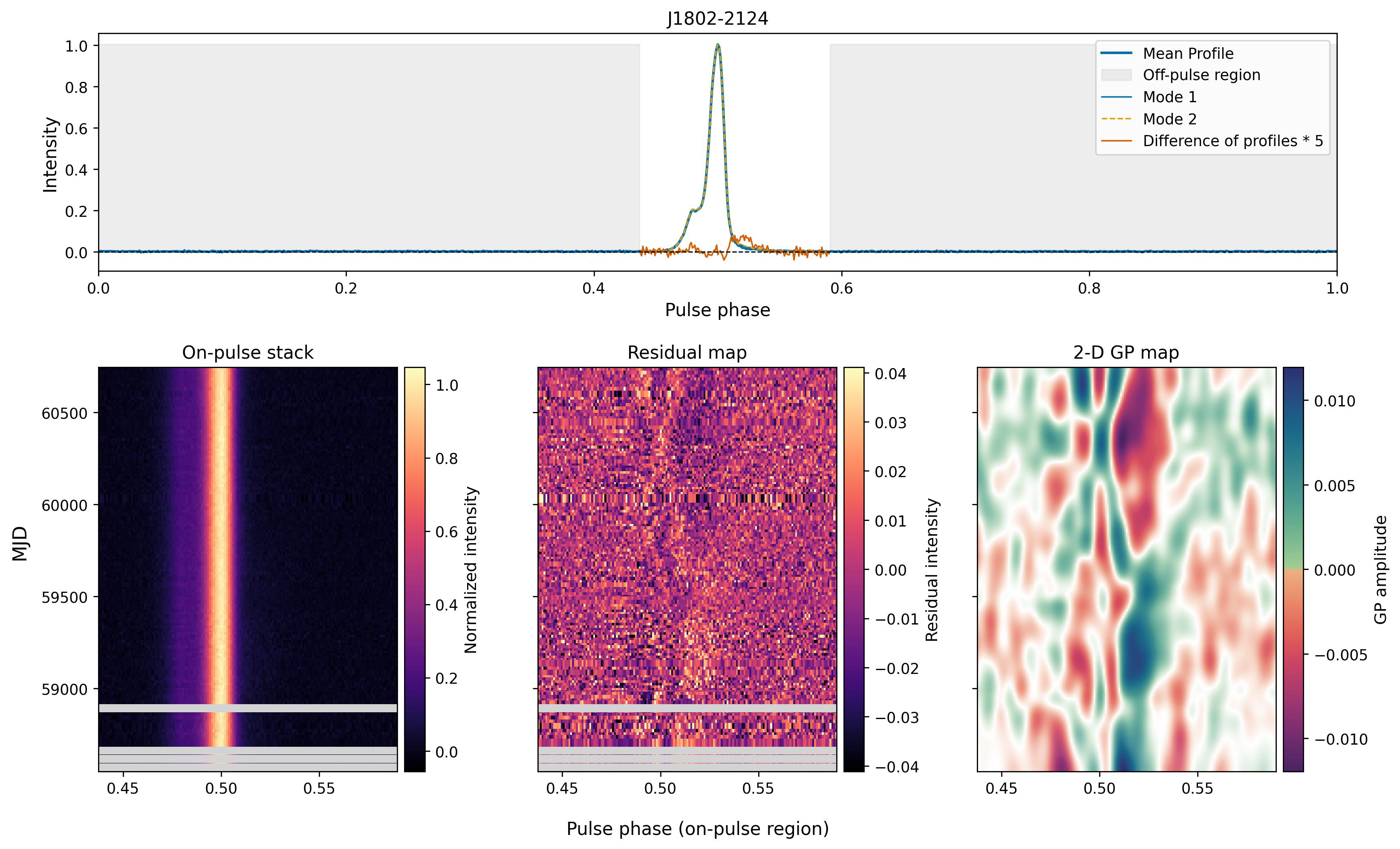}
    \caption{As Figure \ref{fig:J1713} but for PSR~J1802$-$2124. The variability consists of weak stochastic intensity fluctuations concentrated around the peak of the pulse profile, with the Gaussian-process reconstruction revealing coherent temporal correlations that are only marginally visible in the residual map.}
    \label{fig:J1802}
\end{figure*}

In our analysis, PSR~J1802$-$2124 exhibits subtle but statistically significant pulse-profile variability. The observed changes appear stochastic in nature, with temporally correlated intensity fluctuations occurring over multiple timescales. 
The variability is primarily localised around the main pulse component, where small but coherent changes in profile shape are visible. 
Although the residual map is dominated by noise, the correlated structures become more evident in the two-dimensional Gaussian-process reconstruction shown in Figure~\ref{fig:J1802}, confirming that the observed variability is unlikely to arise purely from white noise.

\subsection{PSR J1804$-$2858}

In our analysis, PSR~J1804$-$2858 exhibits broad pulse-profile variability distributed across multiple pulse phases. Owing to the unusually wide and multi-component nature of its pulse profile, quantifying the variability using simple peak-based metrics is challenging. Instead of a localized change, the variability appears as low-amplitude but temporally correlated shape evolution across a broad phase range.

The residual map shows distributed fluctuations across the profile, while the two-dimensional Gaussian-process reconstruction reveals coherent large-scale structures, particularly around the central broad emission component, indicating that the observed changes are not consistent with purely stochastic noise. The variability appears to be dominated by gradual redistribution of intensity between neighbouring profile components rather than abrupt mode switching or isolated transient events.
\begin{figure*}
    \centering
    \includegraphics[width=0.86\linewidth]{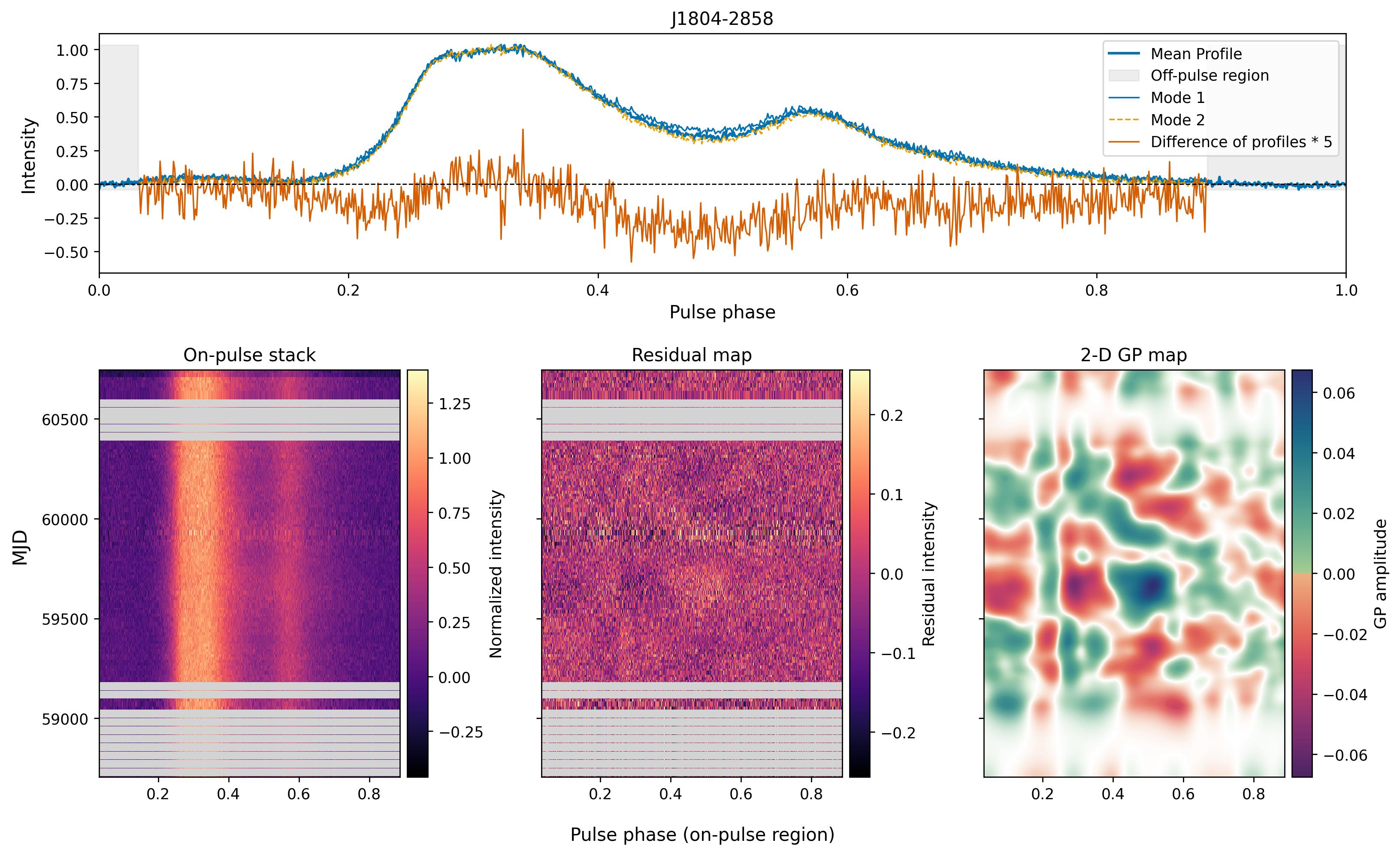}
    \caption{As Figure \ref{fig:J1713} but for PSR~J1804$-$2858. Owing to the broad pulse profile, the observed variability is distributed over a wide range of pulse phase, producing complex stochastic changes that are recovered as coherent structures in the Gaussian-process reconstruction.}
    \label{fig:J1804}
\end{figure*}
\section{Other probable intrinsically variable pulsars}
\label{sec:other_intr_pulsar}

In addition to the pulsars discussed in Section~\ref{sec:Variable_Pulsar}, we identified a small number of pulsars that show tentative evidence for intrinsic pulse-profile variability. These sources exhibit coherent temporal changes in the 2-D Gaussian-process reconstructions and associated principal-component time series, but the amplitudes of the profile variations are sufficiently small that they cannot be classified as secure detections with the current dataset. In several cases, characteristic timescales can be identified; however, the profile evolution remains close to the noise level and lacks the clear, long-lived structures seen in the confirmed intrinsically variable pulsars. We therefore classify these objects as candidate intrinsically variable pulsars and include them here for completeness. Future observations with higher sensitivity, longer observing baselines, and denser cadence will be required to confirm the nature of these profile variations.

\subsection{PSR J1600$-$3053}

\begin{figure*}
    \centering
    \includegraphics[width=0.86\linewidth]{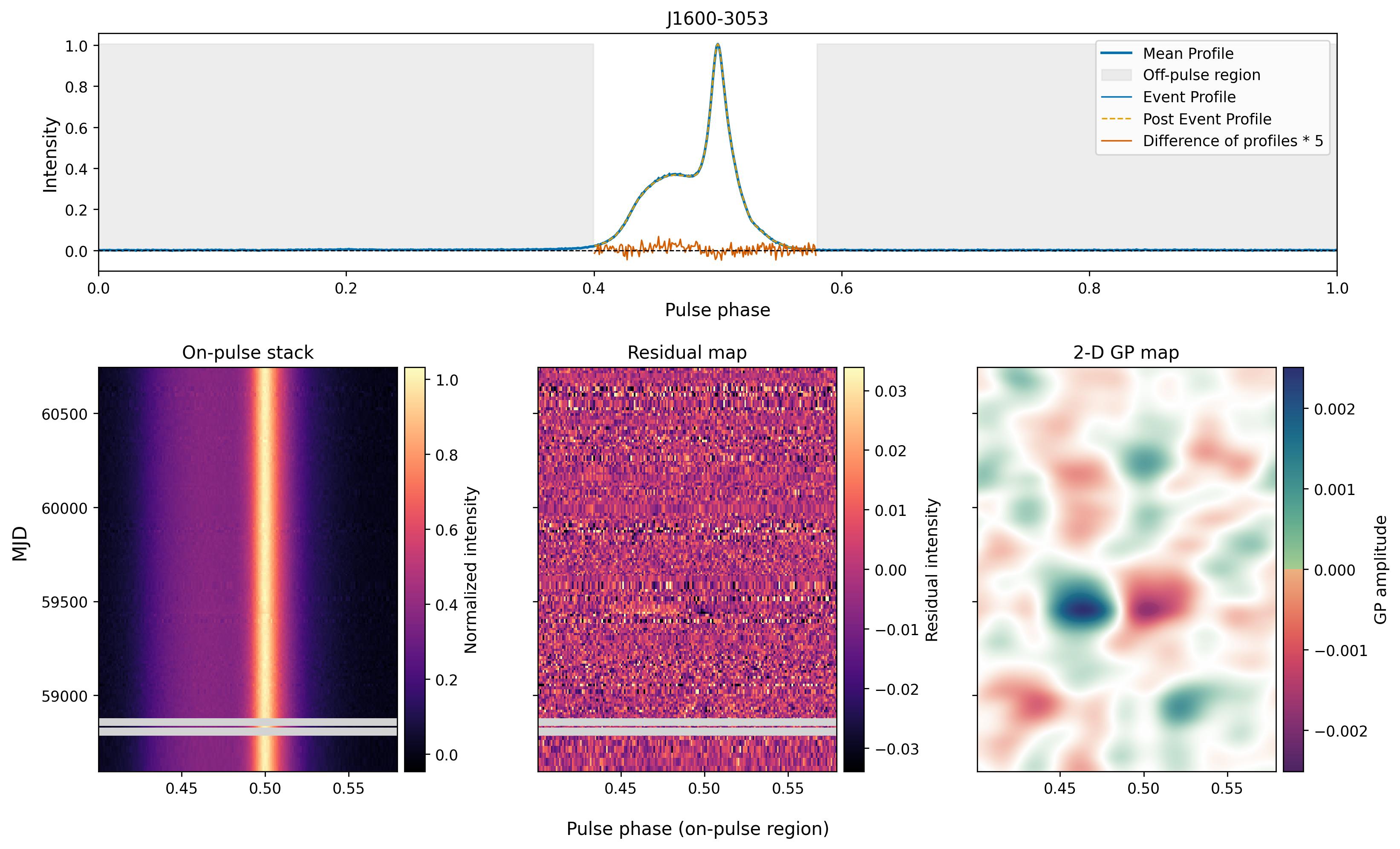}
    \caption{As Figure \ref{fig:J1713} but for PSR~J1600$-$3053. Although the profile variation is only $\sim1.5$ per cent of the pulse amplitude and is barely distinguishable in the raw residual map, the Gaussian-process model recovers a coherent, temporally correlated feature, demonstrating its ability to identify subtle profile variations close to the noise level.}
    \label{fig:J1600}
\end{figure*}

The amplitude of the variation is small ($\sim$1.5 per cent of the pulse profile), making it difficult to distinguish directly from the stochastic noise present in the stacked residual map. Nevertheless, the 2-D Gaussian-process reconstruction successfully isolates a coherent, temporally correlated feature, indicating that the observed variation is suggestive of a temporally correlated signal rather than purely white noise. The profile change is confined primarily to the leading shoulder of the main pulse as shown in Figure~\ref{fig:J1600}, with no evidence for long-term secular evolution or quasi-periodic behaviour.
\subsection{PSR J1603$-$7202}

In our analysis, this pulsar exhibits weak profile variability distributed throughout the observing span, with no evidence for a distinct one-off event or long-term secular evolution. Instead, the variations appear as stochastic, temporally correlated changes that are primarily confined to the two main pulse components. Although the amplitude of the profile changes is small, the 2-D Gaussian-process reconstruction recovers coherent structures that are not readily apparent in the residual map, suggesting the presence of low-level intrinsic profile variability.
\begin{figure*}
    \centering
    \includegraphics[width=0.86\linewidth]{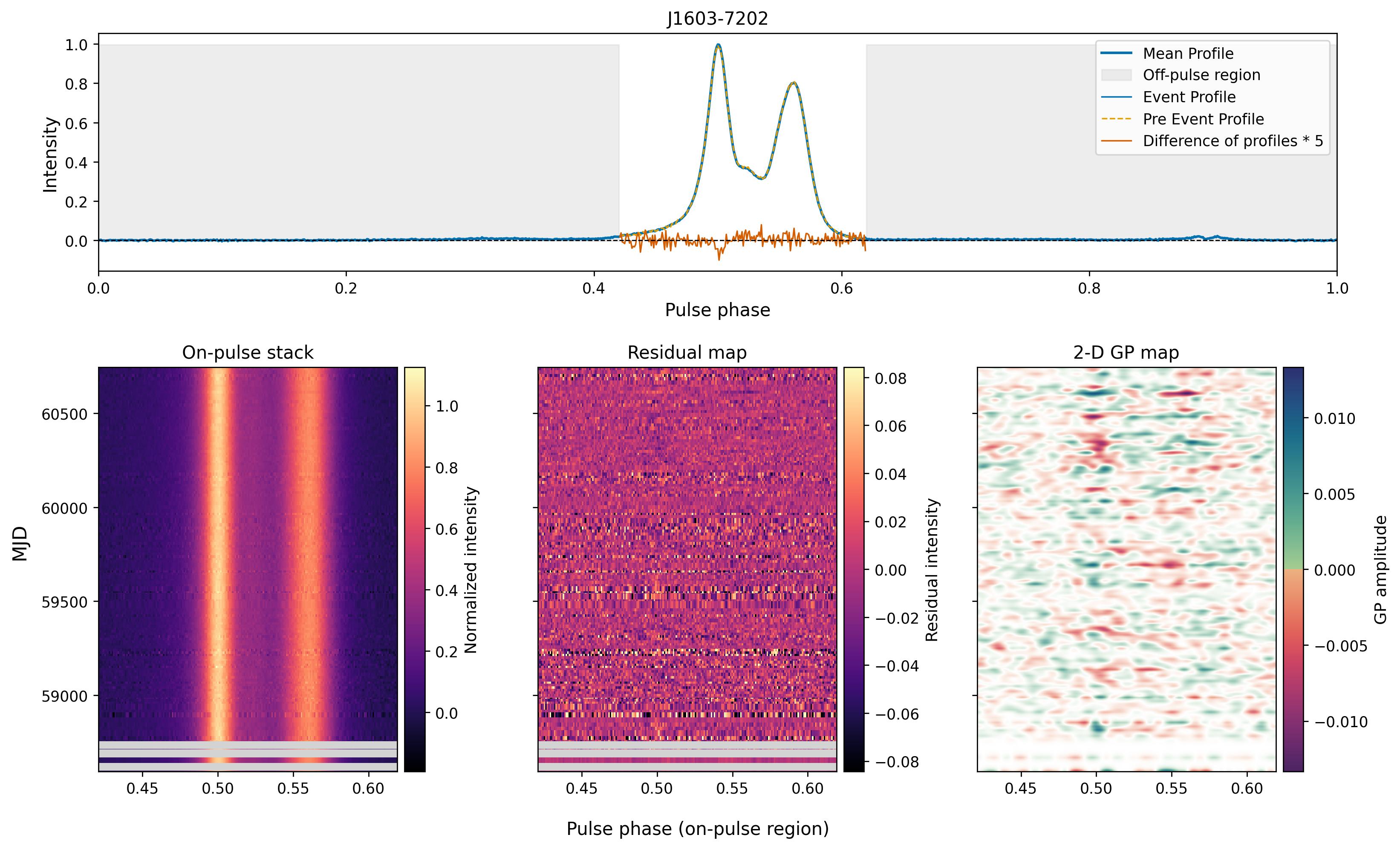}
    \caption{As Figure \ref{fig:J1713} but for PSR~J1603$-$7202. Unlike pulsars exhibiting isolated profile-change events, PSR~J1603$-$7202 displays weak, stochastic, temporally correlated variations distributed across the full observing span, predominantly affecting the two main pulse components.}
    \label{fig:J1603}
\end{figure*}

\subsection{PSR J1327$-$0755}
\begin{figure*}
    \centering
    \includegraphics[width=0.86\linewidth]{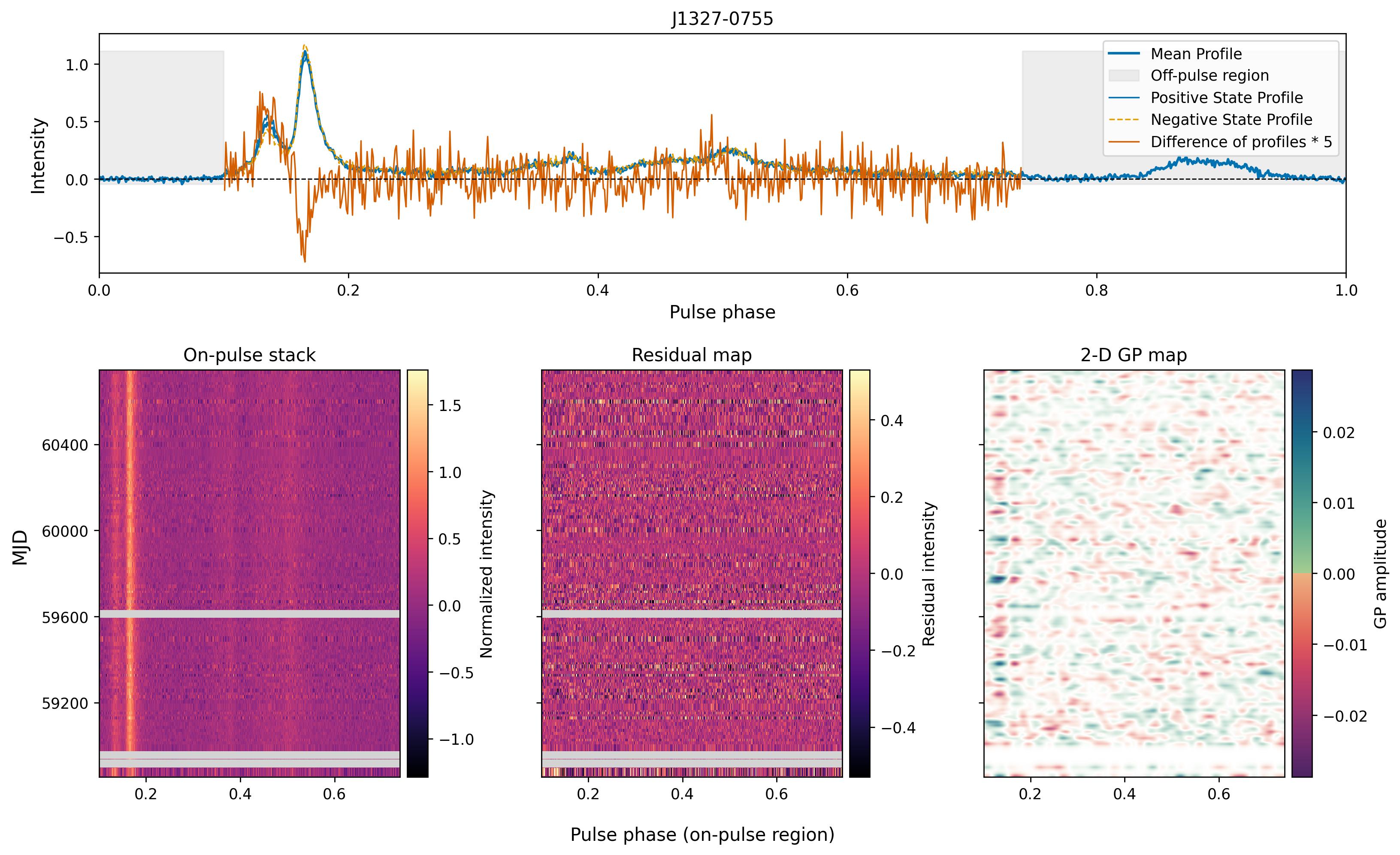}
    \caption{As Figure~\ref{fig:J1713} but for PSR J1327$-$0755. The pulsar exhibits weak, low-amplitude profile variations that are close to the noise level, with no clearly defined long-lived mode transition.}
    \label{fig:J1327}
\end{figure*}

PSR~J1327$-$0755 exhibits weak evidence for long-term pulse-profile variability. The differences between the two representative profile states are concentrated primarily in the leading component of the pulse profile, with smaller changes distributed across the remainder of the on-pulse region (Figure~\ref{fig:J1327}). Although the residual map and 2-D Gaussian-process reconstruction are dominated by low-amplitude fluctuations, a Lomb--Scargle analysis of the principal-component time series yields a characteristic timescale of $83\pm5$~d. This suggests that the profile variability may contain a weak quasi-periodic component. However, because the amplitude of the variation is comparable to the noise level and no clearly defined repeating structure is visible in the profile maps, we classify PSR~J1327$-$0755 as a candidate intrinsically variable pulsar rather than a secure detection.


\bsp	
\label{lastpage}
\end{document}